\documentclass[longauth]{aa}
\usepackage[varg]{txfonts}
\usepackage{graphicx}
\usepackage{float}
\usepackage{natbib}
\bibpunct{(}{)}{;}{a}{}{,}
\usepackage[colorlinks,linkcolor=blue,urlcolor=blue,citecolor=blue]{hyperref}
\usepackage{threeparttablex}
\usepackage{longtable}
\usepackage{lscape}
\usepackage{bm}
\usepackage{verbatim}
\usepackage{makecell}
\usepackage{caption}
\usepackage{subcaption}
\usepackage{multirow}
\usepackage{booktabs}
\usepackage{threeparttable}
\usepackage[dvipsnames]{xcolor}
\usepackage{colortbl}
\usepackage{array}
\usepackage{tabularx}
\usepackage{siunitx}

\definecolor{red}{RGB}{255,0,0}
\definecolor{brown}{RGB}{139,69,19}
\newcommand{\modot}{\ensuremath{\mathrm{M}_{\odot}}}

\newcommand{\mum}{\ensuremath{\mu\mathrm{m}}}
\makeatletter
\g@addto@macro\bfseries{\boldmath}
\makeatother

\begin{document}

\title{Star formation quenching precedes morphological transformation in COSMOS-Web's richest galaxy groups}
\titlerunning{Quenching prior to morphological change in COSMOS-Web groups}

\author{Z.~Ghaffari\thanks{email: zohreh.ghaffari@inaf.it}\inst{1,2}
    \and G.~Gozaliasl\inst{3,4}
    \and A.~Biviano\inst{1,2}
    \and G.~Toni\inst{5,6,7}
    \and S.~Taamoli\inst{9}
    \and M.~Maturi\inst{7,8}
    \and L.~Moscardini\inst{5,6,11}
    \and A.~Zacchei\inst{1,2}
    \and F.~Gentile\inst{12,6}
    \and M.~Haas\inst{13}
    \and H.~Akins\inst{14}
    \and R.~C.~Arango-Toro\inst{15}
    \and Y.~Cheng\inst{26}
    \and C.~Casey\inst{16,14,17}
    \and M.~Franco\inst{12,14}
    \and S.~Harish\inst{10}
    \and H.~Hatamnia\inst{9}
    \and O.~Ilbert\inst{15}
    \and J.~Kartaltepe\inst{10}
    \and A.~H.~Khostovan\inst{18,10}
    \and A. M.~Koekemoer\inst{19}
    \and D.~Liu\inst{21}
    \and G. A. Mamon\inst{20}
    \and H.~J.~McCracken\inst{20}
    \and J.~McKinney\inst{14}
    \and J.~Rhodes\inst{22}
    \and B.~Robertson\inst{23}
    \and M.~Shuntov\inst{17,24,25}
    \and L.~Yang\inst{10}
}
\institute{
    INAF Osservatorio Astronomico di Trieste, Via G.B. Tiepolo 11 e via Bazzoni 2. 34100, Trieste, Italy \label{1}
    \and IFPU: Institute for Fundamental Physics of the Universe, Via Beirut, 2, 34151, Italy \label{2}
    \and Department of Computer Science, Aalto University, PO Box 15400, Espoo, FI-00076, Finland \label{3}
    \and Department of Physics, University of Helsinki, P. O. Box 64, FI-00014 Helsinki, Finland \label{4}
    \and University of Bologna - Department of Physics and Astronomy ``Augusto Righi'' (DIFA), Via Gobetti 93/2, I-40129 Bologna, Italy \label{5}
    \and INAF - Osservatorio di Astrofisica e Scienza dello Spazio di Bologna, via Gobetti 93/3, 40129 Bologna, Italy \label{6}
    \and Zentrum für Astronomie, Universität Heidelberg, Philosophenweg 12, 69120 Heidelberg, Germany \label{7}
    \and ITP, Universität Heidelberg, Philosophenweg 16, 69120, Heidelberg, Germany \label{8}
    \and Department of Physics and Astronomy, University of California, Riverside, 900 University Avenue, Riverside, CA 92521, USA \label{9}
    \and Laboratory for Multiwavelength Astrophysics, School of Physics and Astronomy, Rochester Institute of Technology, 84 Lomb Memorial Drive, Rochester, NY 14623, USA \label{10}
    \and INFN -- Sezione di Bologna, Viale Berti Pichat 6/2, 40127, Bologna, Italy \label{11}
    \and CEA, Université Paris-Saclay, Université Paris Cité, CNRS, AIM, 91191, Gif-sur-Yvette, France \label{12}
    \and Ruhr University Bochum, Faculty of Physics and Astronomy, Astronomical Institute (AIRUB), 44780 Bochum, Germany \label{13}
    \and The University of Texas at Austin, 2515 Speedway Blvd Stop C1400, Austin, TX 78712, USA \label{14}
    \and Aix Marseille Univ, CNRS, LAM, Laboratoire d'Astrophysique de Marseille, Marseille, France \label{15}
    \and Department of Physics, University of California, Santa Barbara, Santa Barbara, CA 93106 USA \label{16}
    \and Cosmic Dawn Center (DAWN), Denmark \label{17}
    \and Department of Physics and Astronomy, University of Kentucky, 505 Rose Street, Lexington, KY 40506, USA \label{18}
    \and Space Telescope Science Institute, 3700 San Martin Drive, Baltimore, MD 21218, USA \label{19}
    \and Institut d'Astrophysique de Paris, UMR 7095, CNRS, and Sorbonne Université, 98 bis boulevard Arago, 75014 Paris, France \label{20}
    \and Purple Mountain Observatory, Chinese Academy of Sciences, 10 Yuanhua Road, Nanjing 210023, China \label{21}
    \and Caltech/IPAC, MS 314-6, 1200 E. California Blvd. Pasadena, CA 91125, USA \label{22}
    \and Department of Astronomy and Astrophysics, University of California, Santa Cruz, 1156 High Street, Santa Cruz, CA 95064, USA \label{23}
    \and Niels Bohr Institute, University of Copenhagen, Jagtvej 128, 2200 Copenhagen, Denmark \label{24}
    \and University of Geneva, 24 rue du Général-Dufour, 1211 Genève 4, Switzerland \label{25}
    \and Department of Astronomy, University of Washington, Seattle, WA 98195, USA \label{26}
}
\date{Received September 15, 2025; accepted , 2025}
\abstract{We analyzed the 25 richest galaxy groups in COSMOS-Web across the redshift range $z = 0.18$--3.65, identified using the Adaptive Matched Identifier of Clustered Objects (AMICO) algorithm. The groups have about 20--30 galaxies with a high ($>$75\%) membership probability. Our study reveals both passive-density and active-density relations,
with late-type galaxies (LTGs) preferring both higher central overdensities than early-type galaxies (ETGs) across all groups, and secondly many massive LTGs have colors typical for quiescent galaxies. We identify red sequences (RS) in 5 out of 25 galaxy groups, prominently established at $z < 1$, with early emergence in the RS locus up to $z \sim 2.2$.
This finding suggests that group environments represent a transitional phase where star formation quenching precedes morphological transformation, contrasting with the classical morphology–density relation observed in rich clusters. In the central regions of the identified groups, within a radius of $\sim33\arcsec$ ($100\,\mathrm{kpc}$) from the group centers, we identified 86 galaxies. Among them, 23 ($\sim27\%$) were classified as ETGs and 63 ($\sim73\%$) as LTGs.
High-mass galaxies ($M_\star > 10^{10.5}\,\modot$) undergo rapid, transformative quenching over $\sim 1\,\mathrm{Gyr}$, becoming predominantly spheroidal ETGs, indicating that morphological transformation accelerates dramatically in the most massive systems during the epoch of peak cosmic star formation. Intermediate-mass galaxies ($10^9 < M_\star/\modot < 10^{10.5}$) show mild quenching, while low-mass galaxies ($M_\star < 10^9\,\modot$) remain largely star-forming, with environmental processes gradually suppressing star formation without destroying disk structures, suggesting that environmental quenching in groups operates on longer timescales than mass quenching. Overall, mass-dependent quenching dominates at the high-mass end, while environment-driven quenching shapes lower-mass systems, highlighting the dual nature of galaxy evolution across cosmic time. The fraction of  HLAGN for both groups and field galaxies increases with redshift, peaking at $z\,\sim\,2$, with groups consistently showing a higher AGN fraction than field. We suggest that AGN feedback plays a partial role in the rapid cessation of star formation in high-mass galaxies, while mergers may contribute to triggering AGN activity.
}
\maketitle
\section{Introduction}
The cosmic structure grows in a hierarchical bottom$-$up fashion: small dark matter halos collapse first at high redshift and, through successive mergers and accretion, assemble into the massive halos we observe today \citep{Springel2005}. The evolving halo mass function implies that, in the present Universe, the bulk of the mass density is contained within halos of about $10^{13}\,\modot$, the scale of galaxy groups. These groups therefore act as the fundamental building blocks of large-scale structure (LSS) and as laboratories in which environmental processes govern galaxy evolution. The properties of galaxies remain tightly coupled to their surrounding dark-matter halos, particularly the most massive systems that dominate their local potential wells \citep{Navarro1997, Dekel2009}. Findings from prolonged studies indicate that fossil-progenitor luminosity functions and evolving magnitude gaps already encode accelerated assembly, underscoring the dynamical maturity of rich environments \citep{Gozaliasl2014b,Gozaliasl2014a}.

The connection between star formation activity and environment is a key aspect of galaxy evolution \citep{Alberts2022}, with observational work across the local Universe reinforcing this picture. In the local Universe, star formation rates (SFRs) correlate with both galaxy stellar mass and local density: massive galaxies in dense environments are typically red and passive, while lower-mass galaxies retain star formation \citep{Kauffmann2004,Baldry2006,Boselli2006,Peng2010,Peng2012,Kawinwanichakij2017}. Studies of the Coma Supercluster show systematic changes in galaxy properties from voids to filaments, groups, and clusters, suggesting increasingly divergent evolutionary pathways as a function of environment \citep{Cybulski2014, Gavazzi2010, Mahajan2010}.
Recent COSMOS connectivity measurements further demonstrate that the location of groups within the cosmic web modulates their mass-assembly histories, strengthening the link between environment and galaxy evolution \citep{Darragh2019,Taamoli2024}.

In dense environments such as galaxy clusters and groups, galaxies exhibit suppressed SFRs, leading to a higher fraction of red-sequence galaxies and more elliptical morphologies compared to field galaxies \citep{Dressler1980, Postman84, Whitmore93}. This gives rise to well-established correlations, including the SFR--density and morphology--density relations \citep{Oemler1974, Dressler1980}, linking local galaxy density to typical galaxy properties. Studies using large samples, such as the Galaxy Zoo project \citep{Bamford2009}, show that galaxy color responds more rapidly to environment than morphology: high-mass galaxies are mostly red across all environments, while low-mass galaxies transition from blue in low-density regions to red in dense regions. This indicates that environmental factors affect galaxy color more quickly than morphological changes, highlighting the distinct color--density and morphology--density relations \citep{Crossett2013}.
Interestingly, at $z \sim 1$, this relation has been reported to reverse, with galaxies in denser regions exhibiting higher SFRs rather than lower \citep{Cucciati2006, Elbaz07, Patel2009}, although the persistence of this reversal remains a subject of debate in subsequent studies \citep{Grutzbauch2011, Popesso2011, Ziparo2014}. This reversal challenges models predicting such behavior only at earlier epochs ($z>2$) and lower intensity \citep{Sazonova2020, Mei2023}, indicating that enhanced star formation in dense environments is not solely merger-driven.

Environment-driven quenching also evolves with redshift. At $z<1$, environmental quenching dominates in clusters and groups, with higher environmental quenching efficiencies (EQEs) in massive halos and cluster cores, while lower-mass galaxies and outskirts retain star formation \citep{Wetzel2012, Muzzin2013, Balogh2014, Balogh2016, vanderburg2018, Pintos-Castro2019, Reeves2021, Sarron2021}. EQE depends on halo mass, cluster-centric distance, and sometimes stellar mass \citep{Papovich2018, Werner2022, Kawinwanichakij2017}. At higher redshifts ($z \sim 1$--$2$), quenching becomes more complex. Some clusters host quenched populations, others vigorous star formation \citep{Cooke2016, Grutzbauch2012, Newman2014, Quadri2012, Santos2013, Stalder2013, Strazzullo2013}. Massive galaxies ($\log M_\star/\modot \gtrsim 10.85$) are mostly quenched, while lower-mass systems remain star-forming \citep{Lee-Brown2017, Nantais2017}. Observations indicate rapid EQE increase over $\lesssim 1\,\mathrm{Gyr}$ in intermediate-redshift clusters ($z\sim1.3$), consistent with a ``delayed-then-rapid'' quenching scenario \citep{Wetzel2013, Foltz2018, Baxter2023}. Cluster-to-cluster variation is intrinsic rather than due to selection effects \citep{Strazzullo2019}.
Detailed investigations of brightest group galaxies (BGGs) across COSMOS strengthen this picture, charting their stellar masses, star-formation histories, stellar ages, and dynamical states as they migrate through quenching channels \citep{Gozaliasl2016,Gozaliasl2018,Gozaliasl2019,Gozaliasl2020,Gozaliasl2024, Gozaliasl2025}.

Physical processes driving quenching include external mechanisms—tidal interactions, mergers, harassment, and gas stripping (e.g., ram pressure)—and internal mechanisms such as AGN feedback \citep{Gunn1972, Larson1980, Dressler1980, Moore1996, Poggianti1999, Boselli2006, Peng2010, Jaffe2015, Cortese2021, Boselli2022}. At $z>1$, quenching is mainly due to gas exhaustion with minor gas stripping, while massive satellites experience a $\sim 2\,\mathrm{Gyr}$ delay before rapid SFR decline; lower-mass satellites quench more gradually \citep{Balogh2016, Foltz2018, Baxter2023}. By $z<1$, environmental quenching is more efficient, with pre-processing in groups and starvation dominating for low-mass galaxies \citep{Wetzel2013, Bianconi2018, Olave-Rojas2018, Werner2022, Zabelle2023}.

AGN powered by supermassive black holes (SMBHs) regulate star formation in massive galaxies by heating the hot intragroup/intracluster medium (IGrM/ICM) and suppressing cooling flows \citep{McNamara2007, Fabian2012}. Feedback strongly shapes the gas properties of groups and clusters \citep{Eckert2021}, as outflows and jets inject energy into the surrounding medium, preventing efficient cooling and star formation \citep{Gitti2012, Gaspari2020}. While early studies of groups lagged behind clusters due to lower X-ray surface brightness \citep{McNamara2007}, more recent work has revealed IGrM cavities and quantified jet powers of $10^{41}$--$10^{44}\,\mathrm{erg\,s^{-1}}$ \citep{Cavagnolo2010, Dong2010, Shin2016, O'Sullivan2017, Kolokythas2019}, with roughly one third of X-ray luminous groups hosting central AGN activity \citep{O'Sullivan2017}. Radio surveys further confirm that group and cluster centers preferentially host AGN, with central galaxies twice as likely as non-centrals to host radio-mode AGN out to $z>1$ \citep{Best2007, Lin2007, Smol2011, Dunn2006, Kolokythas2019}. Feedback in groups is less efficient than in clusters, avoiding over-heating while still capable of displacing gas from cores \citep{Best2007, Giodini2009}. AGN can either suppress star formation by removing or heating cold gas, or briefly trigger it via gas compression, caused by AGN-driven outflows or jets (e.g. \citet{Sijacki2007}). They follow a self-regulated cycle in which IGrM condensation triggers recurrent jets and outflows, stabilizing the system over Gyr timescales. Thus, the thermal evolution of the IGrM is closely linked to the intertwined feeding and feedback processes \citep{Gaspari2016, Gaspari2020}. Consequently, quantifying the number and spatial distribution of AGN and radio galaxies is a major line of research, revealing that group and cluster centers are special locations for AGN \citep{Best2005, Scoville07, Falder2010, Smol2011, Wylezalek2013, Ghaffari2017, Magliocchetti2018, Ghaffari2021, Vardoulaki2024}. The relative importance of these processes continues to be explored through both observations and simulations \citep{Heckman2014, Harrison2017, Sabater2019}.
A puzzling result by ~\citet{Ghaffari2017, Ghaffari2021} is that for at least 50\% of radio-loud AGN (RLAGN) at $z>1$,
such as the 3C radio galaxies and quasars, a clear projected spatial overdensity
of red galaxies has been found, typical for a galaxy group
(see also ~\citet{Wylezalek2013} for other RLAGN samples);
however, red sequences (RS) essentially escaped detection or appear blurred
in color-magnitude diagrams (CMDs) despite small photometric errors
(~\citet{Ghaffari2021}, see also ~\citet{Kotyla2016}
for a detailed RS investigation). This suggests that these high-$z$ clusters
or groups of galaxies have already assembled (or are in the assembling phase), but their members even when showing red colors typical for quiescent galaxies still have active star formation strong enough to disturb a clear red sequence
as identified in clusters at lower redshift ($z<1$),
see ~\citet{Gladders2000}. Even for COSMOS-Web groups,
the red sequences identified so far show a rather large dispersion
(see ~\citet{Toni2025b}). While the morphological distinction of the high-$z$ 3C sources into ETGs and LTGs by ~\citet{Kotyla2016} relied on HST snapshot images, COSMOS-Web provides superb data.
With the in-depth multi-filter data from James Web Space Telescope (JWST), we can now enhance the typical UVJ color selection of quenched galaxies by incorporating morphological measurements obtained through multiple methods. This combined approach offers a promising way to shed light on the evolution of ETGs and LTGs in dense environments.

The plan of this paper is as follows. We present our data and the selection of galaxy groups in Sects.~\ref{sec:data} and~\ref{sec:group_selection}, respectively, while the characterization of the group environments is discussed in Sect.~\ref{sec:group_environment}. The AGN fraction and its potential impact on massive galaxy quenching are examined in Sect.~\ref{sec:RSD}. The predicted properties of the red sequence are described in Sect.~\ref{sec:red-seq}. The evolution of quenching across different stellar mass bins at both low and high redshift is analyzed in Sect.~\ref{sec:quenching}. Finally, we summarize our findings in Sect.~\ref{sec:summary}.
Throughout this work, we adopt a cosmology with $H_0 = 70\,\mathrm{km\,s^{-1}\,Mpc^{-1}}$, $\Omega_{\mathrm{m}} = 0.3$, and $\Omega_{\Lambda} = 0.7$.

\section{Data}\label{sec:data}
The Cosmic Evolution Survey (COSMOS) maps galaxy, AGN, and dark matter evolution across $0.5<z<6$ over $2\,\mathrm{deg}^2$ with deep multiwavelength imaging and spectroscopy from X-ray to radio, including high-resolution HST observations, providing unprecedented galaxy samples and reducing cosmic variance \citep{Scoville07, Koekemoer07, Taniguchi07, Finoguenov07, Capak07, Lilly07, Smol07, Caputi08, Kocevski09, McCracken12, Lemaux14, Cucciati14, Casey15, Diener15, Hung16, Hasinger18}.
The public COSMOS photometry spans over 30 bands from UV to infrared, enabling accurate photometric redshifts and galaxy properties for $\sim$1 million sources, consistently derived using the \textsc{Le Phare} code \citep{Capak07, Ilbert06, Ilbert09, Casey15, Laigle16, Liu18}.

\subsection{COSMOS-Web Survey and COSMOS2025 catalog}
COSMOS-Web is a JWST treasury program imaging $0.54\,\mathrm{deg}^2$ with NIRCam in the F115W, F150W, F277W, and F444W filters to 5$\sigma$ depths of $27.5--28.2\,\mathrm{mag}$, and mapping $0.19\,\mathrm{deg}^2$ with MIRI F770W to $25.3--26.0\,\mathrm{mag}$ \citep{Casey23, Franco2025, Harish2025}, providing homogeneous high-resolution coverage for group analyses. The COSMOS2025 catalog delivers photometric redshifts, morphologies, and physical parameters for $>7\times10^5$ galaxies, using hybrid fixed-aperture and profile-fitting photometry across 37 bands (0.3--8\,\mum).
Photometric redshifts reach $\sigma_{\mathrm{MAD}}/(1+z) = 0.012$ with an outlier fraction $\eta < 2\%$ for $m_{\mathrm{F444W}}<28$ out to $z\sim 9$, representing a factor of $\sim$2 improvement over COSMOS2020
at $m_{\mathrm{F444W}}\approx26$, remaining below 0.03 over a broad range of magnitudes, with $\gtrsim80\%$ completeness for $\log(M_\star/\modot)\approx9$ at $z\sim10$ and better than $10^9\,\modot$ for all $z < 4$ \citep{Shuntov2025}.
Stellar masses and star-formation rates adopt CIGALE \citep{Boquien2019} modeling from \citet{Arango-Toro2025}, fitting star-formation histories calibrated with Horizon-AGN using Bruzual \& Charlot templates \citep{Bruzual2003} and fixed dust emission/attenuation \citep{Calzetti2000, Dale2014}, producing consistent $M_\star$–SFR relations without AGN torus components. To capture rest-frame optical morphologies and the 4000\,\AA\ break, JWST photometry is complemented with HST/ACS F814W imaging \citep{Koekemoer07} and Subaru Hyper
Suprime-Cam imaging \citep{Taniguchi07}. COSMOS-Web structural measurements \citep{Yang2025} use the Python package \texttt{Galight} \citep{Ding2020}\footnote{\url{https://github.com/dartoon/galight}}, revealing that star-forming galaxies maintain roughly constant size--mass and surface density relations, while quiescent systems show steeper structural scaling and a compactness threshold at $\log \Sigma_\ast \sim 9.5$--$10~\modot~\mathrm{kpc}^{-2}$.
\subsection{VLA-COSMOS data}
The VLA-COSMOS 3~GHz Large Project \citep{Delvecchio17} supplies deep radio continuum imaging and source classifications out to $z\lesssim6$, enabling us to trace radio-mode AGN associated with COSMOS-Web groups. AGN identifications are based on a three-component SED decomposition \citep{Berta2013, Feltre2012} that compares fits with and without a torus contribution (via Multi-wavelength Analysis of Galaxy Physical Properties: MAGPHYS; \citealp{Cunha2008}) and adopts the AGN solution when the improvement exceeds the 99\% confidence level.
\subsection{COSMOS spectroscopic redshift catalog}
\citet{Khostovan2025} present a compilation of $\sim 488\,000$ spectroscopic redshifts for $266\,000$ galaxies over 20 years in the COSMOS field, covering a wide range of masses and redshifts up to $z\sim8$. The compilation is primarily complete for low- to intermediate-mass star-forming and massive quiescent galaxies at $z<1.25$, and low-mass star-forming galaxies at $z>2$.
\subsection{COSMOS-Web group catalog}\label{sec:group_selection}
COSMOS-Web galaxy groups \citep{Toni24, Toni2025} are identified with the AMICO matched-filter
algorithm \citep{Bellagamba18, Maturi19}, which treats galaxies as a clustered signal atop a field background and assigns probabilistic memberships.
Catalog columns include \texttt{ID}, \texttt{RA}, \texttt{DEC}, redshift (\texttt{Z}),
the detection signal-to-noise (S/N) value (\texttt{SN\_NO\_CLUSTER}) without cluster shot-noise contribution, which
serves as the primary indicator of detection significance and sample purity, the intrinsic richness $\lambda_*$ (\texttt{LAMBDA\_STAR})
defined as the sum of membership probabilities for galaxies within $R_{200}$
with $m \le m^* + 1.5$ (with parameters given by the detection model), and the probabilities of being associated with a
group or with the field (\texttt{ASSOC\_PROB} and \texttt{FIELD\_PROB},
respectively). For further details about these quantities, we refer the reader to \citet{Toni2025}. The group search relies on masks based on Gaia DR2
visibility mask and bright stars ($G < 18$), and residual imaging artifacts
are removed through testing and visual inspection \citep{Toni24, Toni2025}.
and residual imaging artifacts are removed through visual inspection \citep{Toni24, Toni2025}.
Applying these criteria yields a total of 1\,678 detections over $0.45\,\mathrm{deg}^2$, down to a detection threshold of $\texttt{S/N} = 6$.
Throughout the analysis we refer to the AMICO centroid as $(\mathrm{RA}_{\mathrm{DET}},\mathrm{Dec}_{\mathrm{DET}},z_{\mathrm{DET}})$, and use these values when cross-matching photometric and spectroscopic measurements. The effect of the masked area is already taken into account, following the same correction procedure described in \citet{Toni2025}.

We characterize the richest groups in each redshift bin by cross-matching AMICO group members with the COSMOS-Web Master Catalog v3.1.0 \citep{Shuntov2025}, from which we obtain redshifts and colors. Stellar masses and SFRs are then supplemented using the catalogs of \citet{Arango-Toro2025}, and Sérsic indices are taken from \citet{Yang2025}.

\section{Galaxy groups environments}\label{sec:group_environment}
We identified spectroscopic group members by selecting all sources within a projected radius of $800\,\mathrm{kpc}$ from each group center and applying a redshift-dependent cut of $\pm 0.01(1 + z_{\mathrm{CL}})$. Reliable redshifts: secure measurements with \texttt{flag = 3, 4, 13, 14} and moderate-confidence measurements with \texttt{flag = 2, 12} when their confidence level is $80\%$ \footnote{\texttt{flag = 3, 4} correspond to high-confidence galaxy redshifts, \texttt{flag = 13, 14} to secure broad-line AGNs, and \texttt{flag = 2, 12} to moderate-confidence measurements with $\sim 80\%$ reliability.}. For each group we counted secure and total matches, which are listed in Table~\ref{tab:richest_clusters}.

To preserve uniform quality we impose the following selection criteria:
\begin{itemize}
    \item Apply a magnitude cut of $\texttt{MAG\_MODEL\_F444W}<27.2$ (Average fraction removed: 16\% ($\mathrm{min}{\sim}2\%,\, \mathrm{max}{\sim}30\%$
    )) to remain within the $\sim95\%$ completeness limit established in Sect.~\ref{sec:phot_comp} and to exclude sources with poorly constrained photometry.
    \item The best photo-$z$ quality flags ($\text{LP\_warn\_fl} = 0$) are considered, because we are using the AMICO membership catalog, which already accounts for this.
\end{itemize}

\subsection{Morphology of galaxy group members}
We use structural measurements from \cite{Yang2025} to classify galaxies into early-type (ETGs) and late-type (LTGs) for studying their evolution from star-forming to quiescent states. Morphological classification is based on the Sérsic index \citep{Sersic1968}, with galaxies having $n_S<2.5$ classified as LTGs and $n_S\ge2.5$ as ETGs \citep{vanderWel2014}. To ensure consistent rest-frame optical sizes across redshift, we adopt the NIRCam filter closest to $\sim8000$\,\AA, minimizing biases from star formation and dust \citep{Gozaliasl2025}, using F115W ($0.05<z\le0.4$), F150W ($0.4<z\le1.0$), F277W ($1.0<z\le3.0$), and F444W ($3.0<z\le4.0$).

\subsection{Selection of rich groups}
\begin{figure}
    \centering
    \includegraphics[width=0.99\linewidth, clip=true]{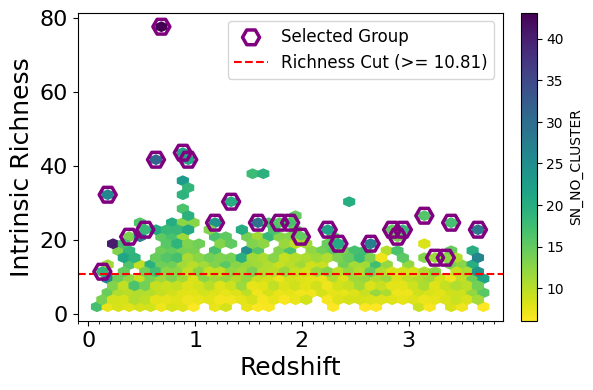}
    \caption{Intrinsic richness $\lambda_\star$ versus redshift (Hexbin map) for the COSMOS-Web AMICO detections. Each bin is colored by the median $\texttt{SN\_NO\_CLUSTER}$, the dashed red line marks the richness threshold $\lambda_\star=10.81$, and the purple outlines highlight the 25 groups retained for detailed analysis.}
    \label{fig:sel-rich}
\end{figure}

Figure \ref{fig:sel-rich} shows a hexbin map of intrinsic richness ($\lambda^{\star}$) versus redshift for the AMICO detections. The color of each hexagonal bin represents the median S/N (\texttt{e.g., SN\_NO\_CLUSTER}) value of the groups it contains. For the selection, the galaxy group sample was binned into 19 redshift intervals, each with a width of $\Delta z=0.2$, spanning $0 < z < 3.8$. Within each redshift bin, we first identified the two groups with the highest intrinsic richness and S/N, yielding a sample of 38 galaxy groups. From these, we further selected 25 groups that display a symmetric annular distribution of photometric redshift–selected galaxy members within a physical scale of $1\,\mathrm{Mpc}$ around the group center $(\mathrm{RA}_{\mathrm{DET}}, \mathrm{Dec}_{\mathrm{DET}})$.
The redshift values for the sample range from 0.18 to 3.65, with S/N values varying from 12.59 to 43.11, and the richness values spanning from 10.81 to 77.57.
The galaxy groups selected for comparison are listed in Table \ref{tab:richest_clusters} and marked with purple hexagons in Fig. \ref{fig:sel-rich}. The increasing trend of intrinsic richness observed up to redshift $z\,\sim\,2$ is expected, as detecting less-rich groups becomes increasingly difficult at higher redshifts.
At $z\,\gtrsim\,2$, numerous low-$\lambda_\star$ detections may be attributed to the fact that AMICO is likely detecting the cores and substructures of clumpy, extended protostructures rather than individual virialized groups.

\subsection{Radial surface density (RSD) in rich groups}\label{sec:RSD}
\begin{figure*}
    \centering
    \begin{subfigure}{0.32\textwidth}
        \centering
        \includegraphics[width=\linewidth, height=4.5cm, clip=true]{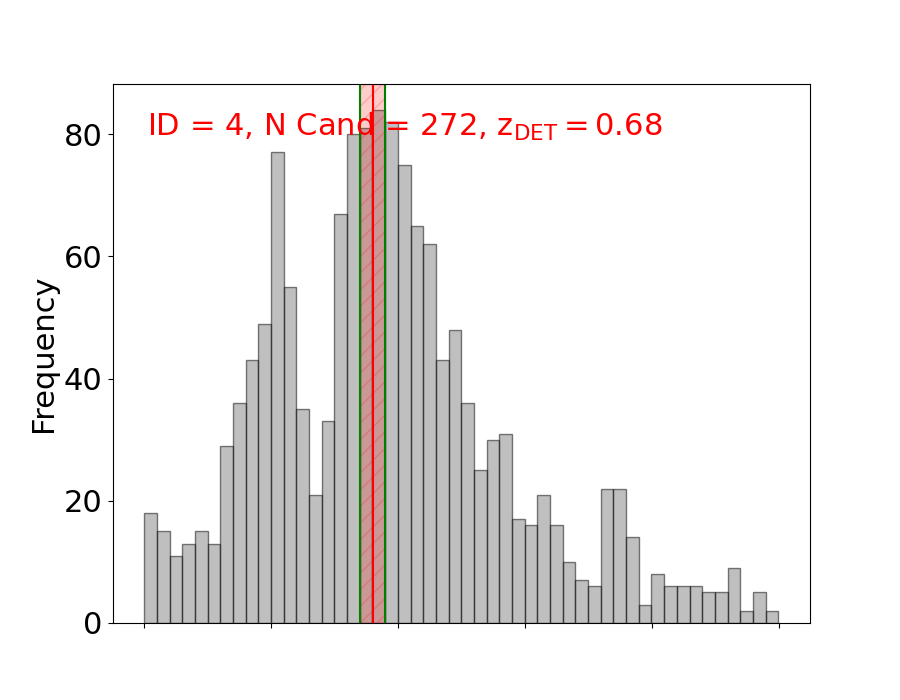}
    \end{subfigure}
    \hfill
    \begin{subfigure}{0.32\textwidth}
        \centering
        \includegraphics[width=0.95\linewidth, height=4.0cm, clip=true]{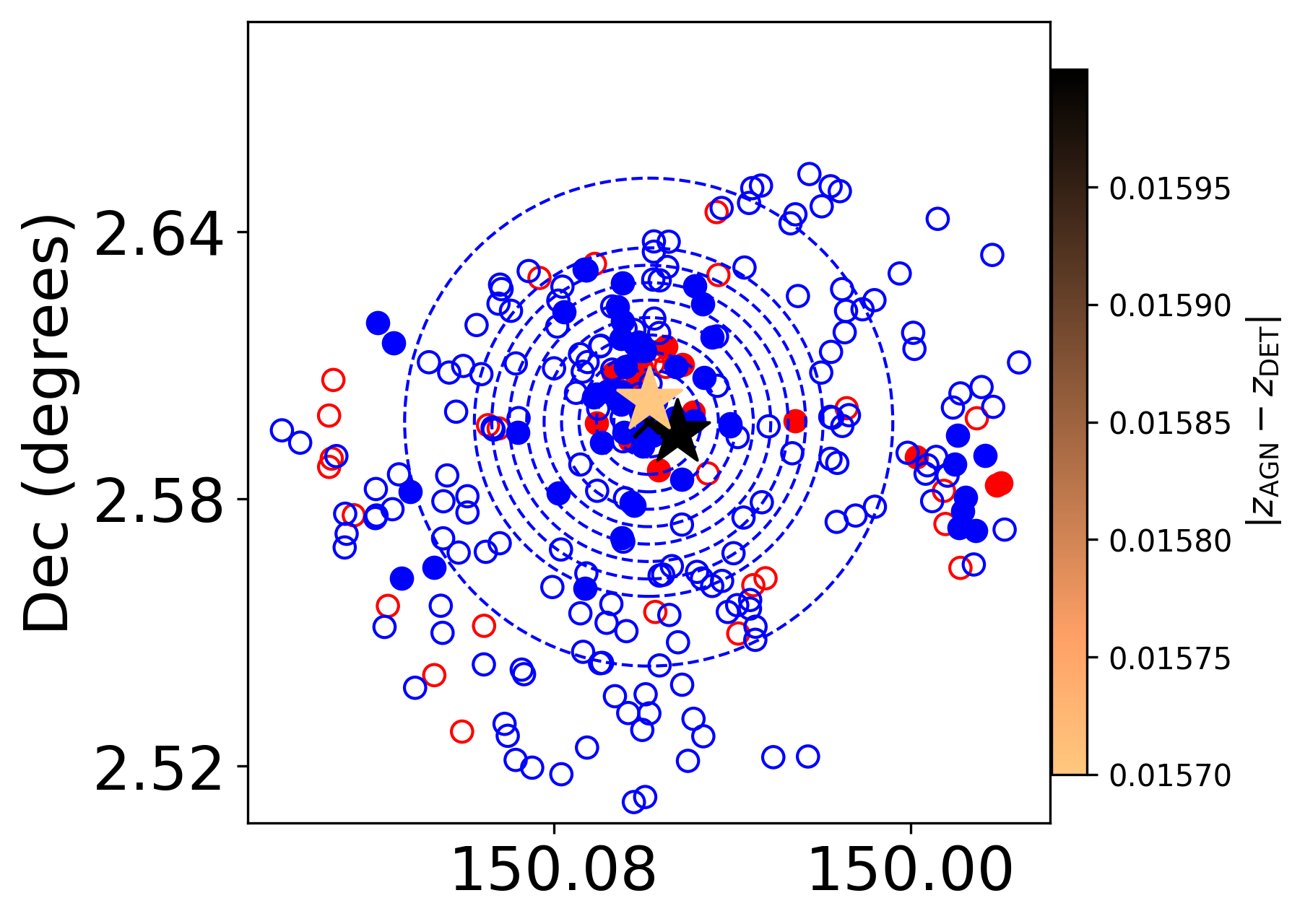}
    \end{subfigure}
    \hfill
    \begin{subfigure}{0.32\textwidth}
        \centering
        \includegraphics[width=\linewidth, height=4.0cm, clip=true]{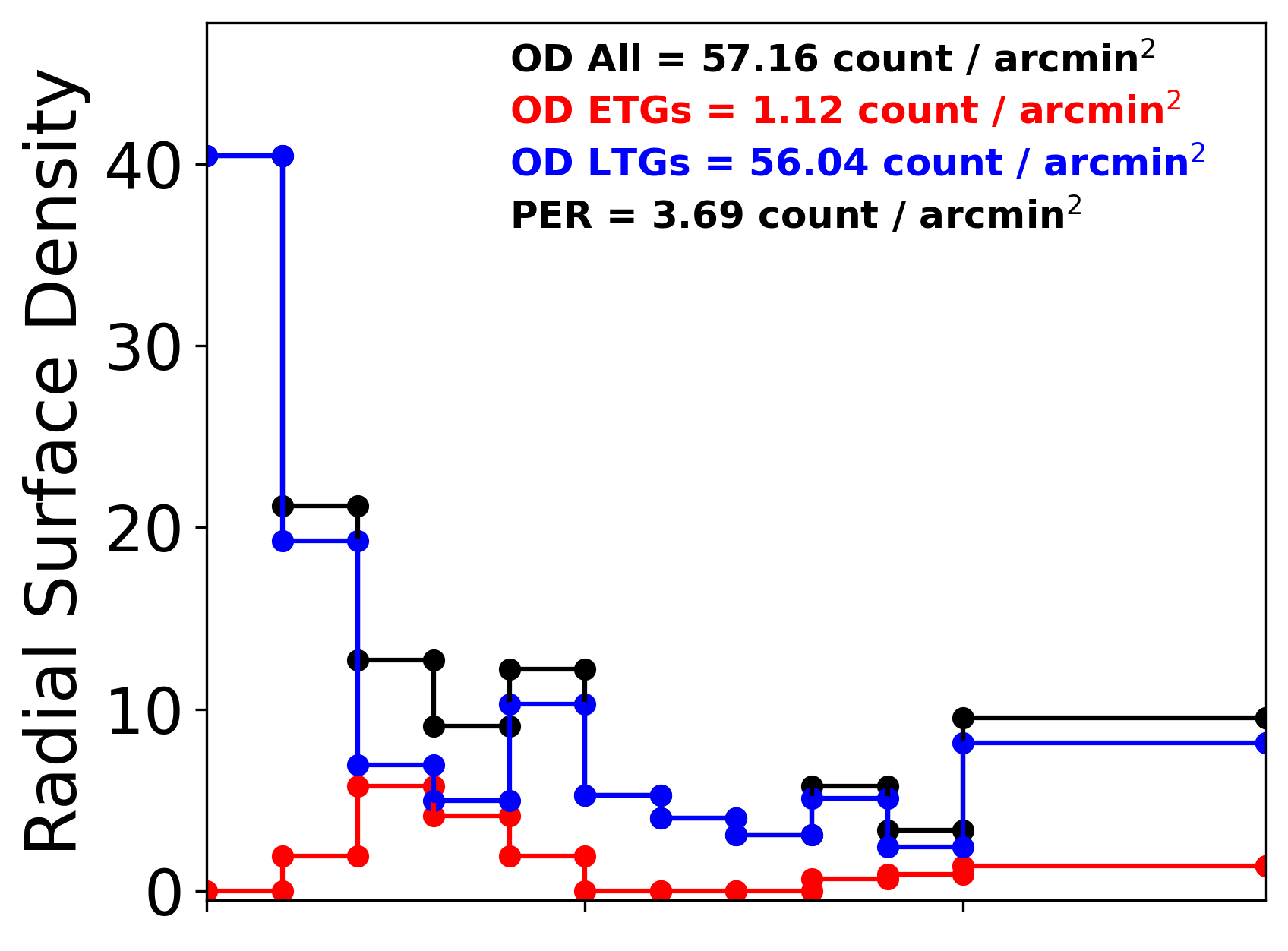}
    \end{subfigure}

    \vspace{-1mm} 

    \begin{subfigure}{0.32\textwidth}
        \centering
        \includegraphics[width=\linewidth, height=4.5cm, clip=true]{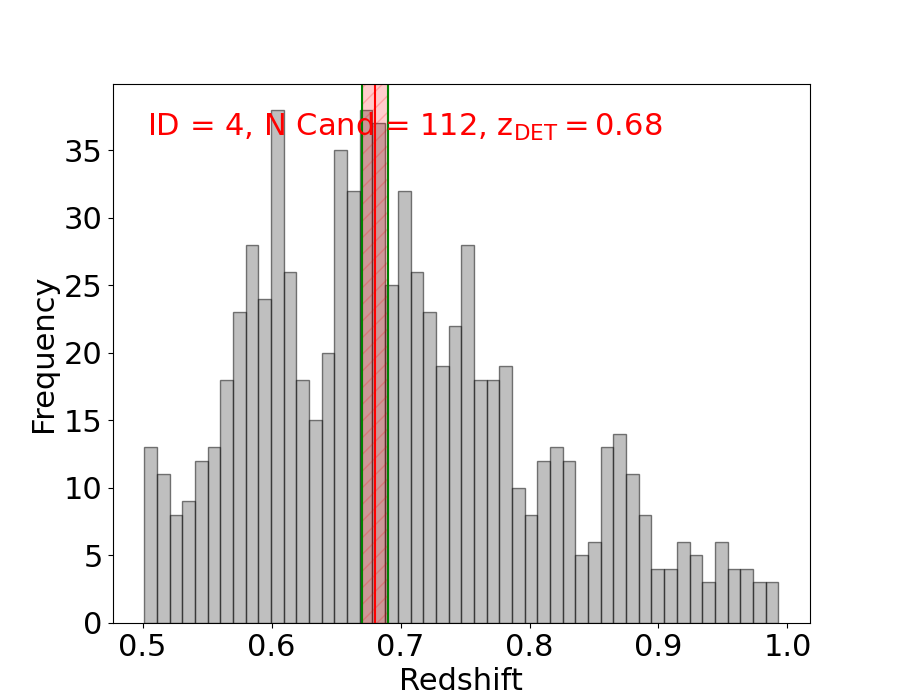}
    \end{subfigure}
    \hfill
    \begin{subfigure}{0.32\textwidth}
        \centering
        \hspace{-1.0cm}\includegraphics[width=0.85\linewidth, height=4.2cm, clip=true]{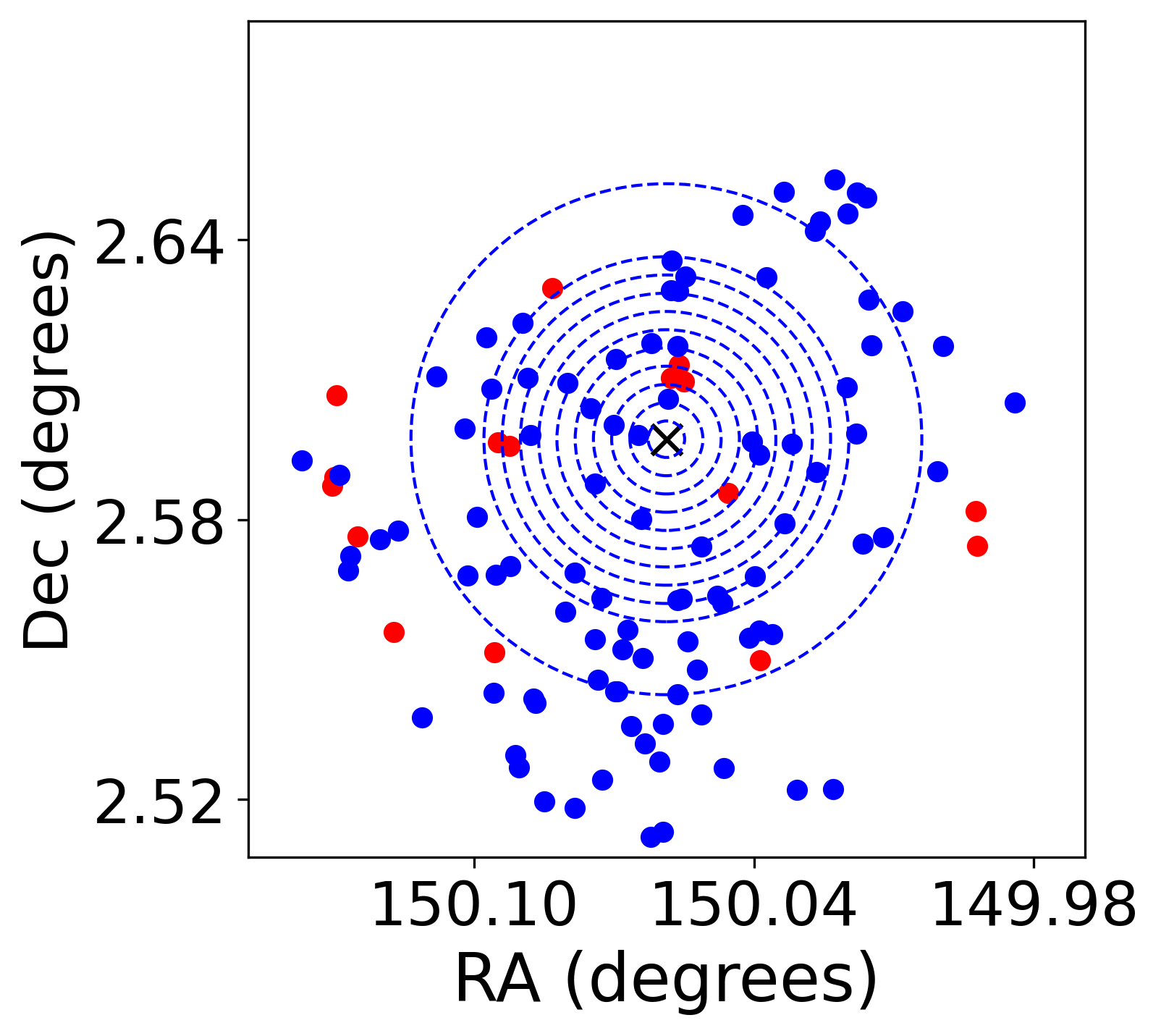}
    \end{subfigure}
    \hfill
    \begin{subfigure}{0.32\textwidth}
        \centering
        \includegraphics[width=\linewidth, height=4.3cm, clip=true]{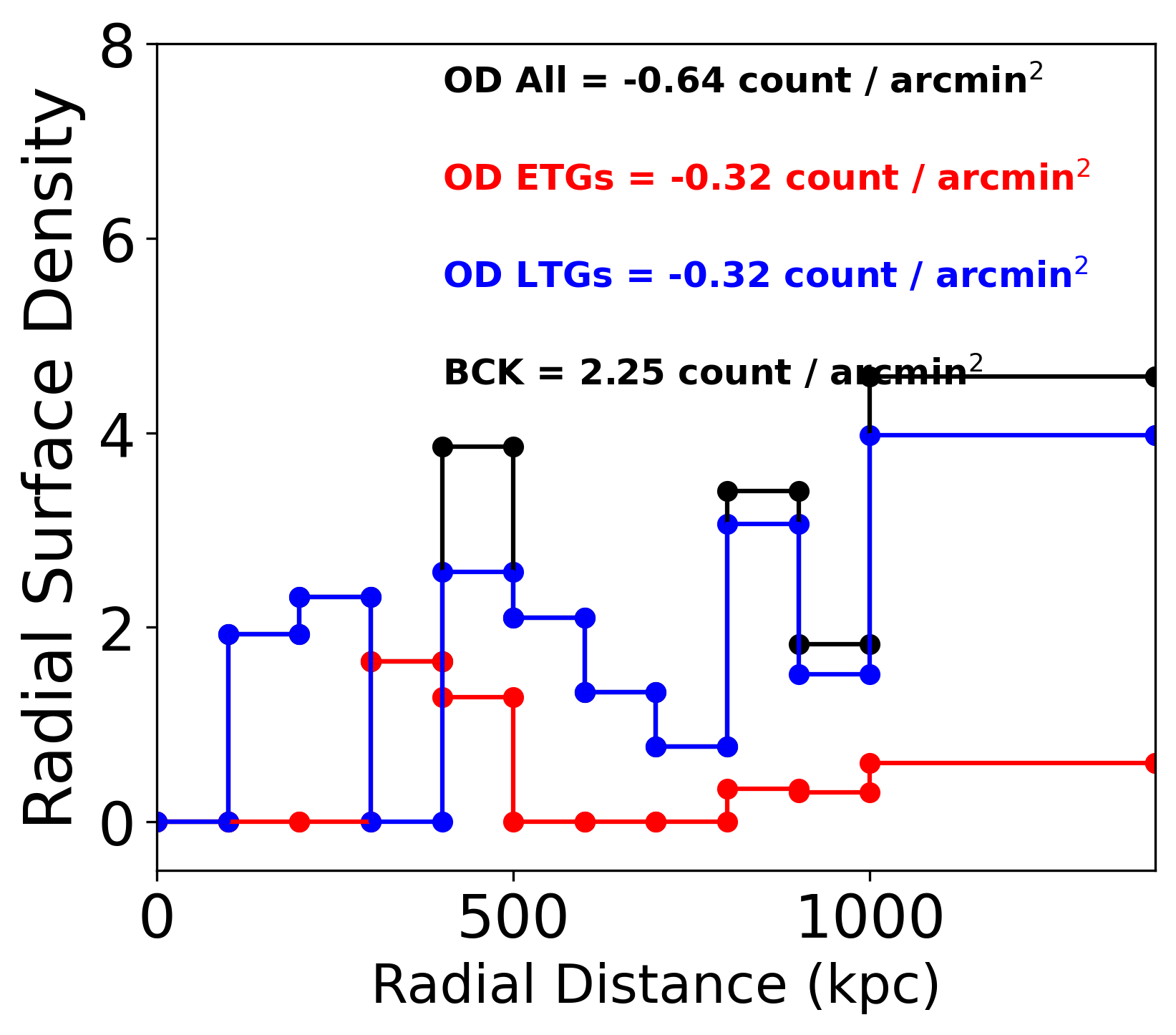}
    \end{subfigure}
\caption{Member-selection diagnostics for group ID~4. Left: redshift distribution of AMICO candidates. The red solid line and hatched region (bounded by green lines) represent $z_{\mathrm{DET}}$ and the adopted selection window $\Delta z = \pm 0.01(1 + z_{\mathrm{DET}})$, respectively. Middle: projected distribution of LTGs (blue) and ETGs (red) in concentric annuli, with galaxies having $\text{assoc\_prob} > 0.5$ shown as filled circles and those with $\text{assoc\_prob} \leq 0.5$ as unfilled circles; matched HLAGN are marked as stars colored by $|z_{\mathrm{AGN}}-z_{\mathrm{DET}}|$. Right: radial surface-density profile showing ETG, LTG, and combined overdensities after subtracting the local background (PER). Panels in the bottom row repeat the analysis for high field-probability galaxies (e.g. FIELD\_PROB > 0.7). Negative values indicate underdensities, where the accumulated density minus the outskirts baseline becomes negative.}
    \label{fig:hist-rsd-annuli}
\end{figure*}

We quantify central overdensities by converting projected separations into physical radii and constructing radial surface density profiles for ETGs and LTGs separately. Each group is divided into concentric $100\,\mathrm{kpc}$ annuli out to $1.4\,\mathrm{Mpc}$ (The choice of $1.4\,\mathrm{Mpc}$ is intended to probe large-scale structure and V-shaped RSD \footnote{V-shape means a secondary rise at larger projected radii, suggesting a nearby overdensity or companion structure.}, rather than the physical size of the groups.), galaxy counts are normalized by annular area (counts\,arcmin$^{-2}$), and the central overdensity (OD) is defined as the density within $r<200\,\mathrm{kpc}$ minus the local background (BCK) measured by the local periphery (PER) between 800 and $1000\,\mathrm{kpc}$. The
average global background is $\sim0.2\,\mathrm{counts\,arcmin^{-2}}$ , whereas local periphery densities reach $\sim1\,\mathrm{counts\,arcmin^{-2}}$ and typically exceed the global value by factors of 3--5; subtracting this local PER minimizes contamination from fore- and background sources in each group. The catalog of COSMOS-Web groups by \citet{Toni2025} lists all potential candidate group member galaxies and, for each galaxy, provides an association probability (\texttt{assoc\_prob}) and a field probability. Many galaxies, however, have relatively low \texttt{assoc\_prob} values ($<0.5$), and galaxies with large photometric redshift uncertainties ($\sigma_z > 0.1$) also enter the catalog. While $\sigma_z$ and \texttt{assoc\_prob} are not simply correlated, the galaxies with largest $\sigma_z$ show the lowest \texttt{assoc\_prob}. In order to select only the most reliable members, we applied a narrow redshift threshold $\pm 0.01(1 + z_{\mathrm{DET}})$ and verified the robustness of each group through the emergence of the red sequence (RS), as discussed in Section~\ref{sec:red-seq}. This approach allows us to assess the reliability of the detected systems and to reduce potential spurious detections and mis-centering effects, which are expected in low-density regions when applying a relatively low S/N threshold of 10.
Fig.~\ref{fig:hist-rsd-annuli} as an example illustrates the member distribution workflow for group ID~4. The left panel shows the redshift distribution of AMICO candidates with the adopted window $\pm0.01(1 + z_{\mathrm{DET}})$ around $z_{\mathrm{DET}}$ (green dashed lines); the middle panel maps projected positions and morphological types (blue LTGs, red ETGs) in concentric annuli, including matched High$-$Luminosity Active Galactic Nuclei (HLAGN) whose redshift offsets are color-coded; and the right panel presents the resulting RSD profile. We rely on spatial and redshift proximity rather than the full AMICO probability distribution to limit spurious associations. We selected galaxy members within a spatial radius of $800~\mathrm{kpc}$ from the catalog's center (RA$_{\rm DET}$ and Dec$_{\rm DET}$) and within a redshift range of $\pm 0.01 \, (1 + z_{\rm DET})$. We then applied the redshift-dependent mass cut described in the appendix Sect.~\ref{sec:stella_mass_completeness} to the remainder of our analysis.
Notice that group ID~304 required a recentering because the catalog' center $(\mathrm{RA}_{\mathrm{DET}},\mathrm{Dec}_{\mathrm{DET}})$ did not trace the dynamical core. We therefore computed a stellar-mass-weighted barycenter, shifting the coordinates by $1.47\arcmin$ east and $2.13\arcmin$ south, which restored the expected central overdensity:
\begin{align*}
\text{RA}_{\mathrm{corrected}}\,=\,
\frac{\sum\limits_{i} \left( \text{RA}_{i} \, M_{i} \right)}
     {\sum\limits_{i} M_{i}}, \text{Dec}_{\mathrm{corrected}}\,=\,\frac{\sum\limits_{i} \left( \text{Dec}_{i} \, M_{i} \right)}
     {\sum\limits_{i} M_{i}}.
\end{align*}
Across all 25 systems, none show ETG overdensities surpassing those of LTGs, indicating that group cores remain dominated by gas-rich disks.
The RSD for ETGs, LTGs, and the combined sample (All) are shown in right panel of Fig.~\ref{fig:hist-rsd-annuli} and Table~\ref{tab:richest_clusters}, with the corresponding calculations (OD, PER, BCK) annotated in the figure. Same plot for a subsample (group IDs: 15, 1, 36, 72, 30, 156) is shown in the appendix (Fig.~\ref{fig:combined_plot_rsd}, while the remaining plots are available online via the Zenodo link: \url{https://zenodo.org/uploads/17407954}).

\subsubsection{Size$-$Redshift relation in rich groups}
\begin{figure}
    \centering
    \hspace*{-1.5cm}\includegraphics[width=0.44\textwidth]{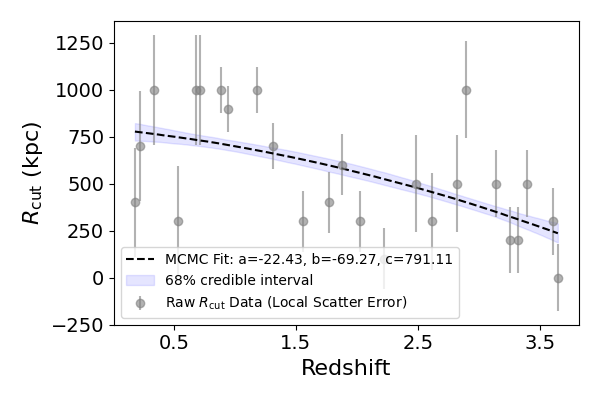}
    \caption{Average radius $R_{\mathrm{cut}}$ as a function of redshift. Blue points show binned means with Poisson errors, and the dashed curve denotes the quadratic fit $R_{\mathrm{cut}}=-22.43 z^2 - 69.27 z + 791.11\,\mathrm{kpc}$.}
    \label{fig:redshift_radius_plot}
\end{figure}

$R_{\mathrm{cut}}$ is the projected physical radius (in kpc) where the group's RSD profile flattens to the background level, representing the effective physical size of the system. Fig.~\ref{fig:redshift_radius_plot} summarizes the RSD-derived $R_{\mathrm{cut}}$ values for the 25 groups. The gray data points represent the individual, with error bars reflecting the local standard deviation in $R_{\mathrm{cut}}$. The robust MCMC fit to this raw dataset reveals a steady decline in characteristic size with increasing redshift (black dashed line). This trend mirrors the mass–observable evolution reported by \citet{Grandis2024} for clusters and is consistent with the expectation that higher-redshift groups are less massive and therefore more compact.

\begin{figure}
    \centering
    \begin{subfigure}{\columnwidth}
        \includegraphics[width=\columnwidth]{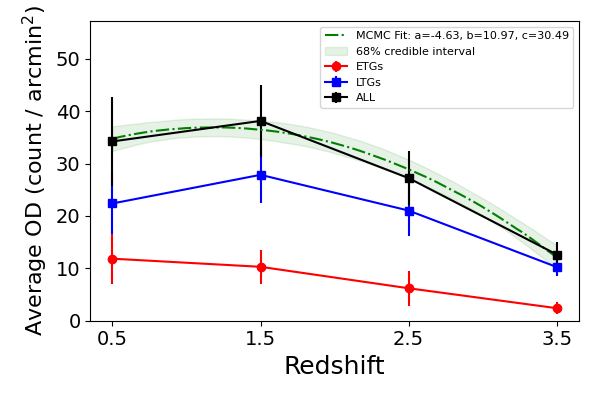}
    \end{subfigure}
    \caption{
    Average overdensity versus redshift for ETGs (red circles) and LTGs (blue squares), and their sum (black), measured within cumulative apertures of $200\,\mathrm{kpc}$. Error bars denote the standard error of the mean, while the dashed curve and shaded regions show the Markov-Chain Monte Carlo (MCMC) fit and its 68\% credible interval for the combined population. Average overdensity of field galaxies is consistent with a constant background level.}
    \label{fig:cum-cod_200}
\end{figure}

\subsubsection{Central regions of rich groups $(<\,100, 200\,\mathrm{kpc})$}\label{sec:central_regions}
Stacking overdensities within a $200\,\mathrm{kpc}$ aperture shows that LTGs maintain higher central densities than ETGs across most redshifts (Fig.~\ref{fig:cum-cod_200}); results for larger apertures (400 and $600\,\mathrm{kpc}$) are presented in Appendix~\ref{fig:cum-cod-200-400}. ETG overdensities decrease from $z \sim 1$ to $z > 2$, while, over the same interval, LTG overdensities show a slight increase, highlighting the continued prevalence of star-forming disks in group cores. Unlike the purely passive-density relation seen in massive clusters \citep{Mei2023}, our groups display concurrent passive- and active-density trends, consistent with gentler potentials that allow LTGs to linger and quench more gradually.
Within $r<100\,\mathrm{kpc}$ we identify 86 members across the 25 groups, of which 63 (73\%) are LTGs and 23 (27\%) are ETGs. The richest cores host up to seven galaxies (IDs~4, 36), whereas one system (ID~14) lacks any central members. Groups with five or more central galaxies exhibit central overdensity (COD) values spanning $\sim 4$ to $\sim 70\,\mathrm{counts\,arcmin^{-2}}$, reflecting the diversity of dynamical states. In several cases the AMICO centroid aligns with an average of multiple density peaks rather than the deepest potential minimum, especially in unrelaxed or lower-mass halos \citep{Toni2025}.

\subsubsection{AGN fraction for all groups}\label{sec:agn_fraction}
\begin{figure}
\centering
\includegraphics[width=0.48\textwidth]{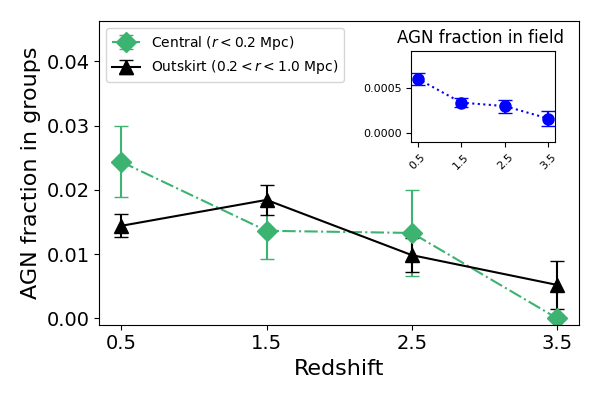}
\caption{Fraction of HLAGNs as a function of redshift for central ($r < 0.2\,\mathrm{Mpc}$) and outskirt ($0.2 < r < 1.0\,\mathrm{Mpc}$) regions. Points show the combined group sample with Poisson uncertainties.}
    \label{fig:agn_fraction_plot}
\end{figure}

We identify HLAGN ($L_{\mathrm{X}} \gtrsim 10^{44}\,\mathrm{erg\,s^{-1}}$) counterparts within $\pm 0.05 \times (1 + z_{\rm DET})$ of each full sample ($\sim1678$) group's redshif\footnote{Five times the selection window adopted for member galaxies, comparable to photometric-redshift uncertainties in wide surveys.} and measure their projected distance from the group center to classify sources as central ($r<0.2\,\mathrm{Mpc}$) or outskirts ($0.2<r<1.0\,\mathrm{Mpc}$). Fractions are computed using only groups that contribute at least one galaxy to the relevant radial bin, ensuring consistent normalization of $N_{\mathrm{AGN}}$, $N_{\mathrm{gal}}$, and $N_{\mathrm{group}}$.
Table~\ref{tab:agn_fraction} and Fig.~\ref{fig:agn_fraction_plot} show that HLAGN are slightly more concentrated in the central regions at $z<1$ (2.4\% versus 1.4\%), while at $z>1$ the outskirts fractions rise to $\sim 2\%$ and exceed or match the central values. AGN activity therefore transitions from being core-dominated at low redshift to more broadly distributed at higher redshift, consistent with an increased supply of cold gas in the outer group environment near cosmic noon.

\begin{table*}
\centering
\caption{Average AGN fraction for COSMOS-Web groups in central and outskirt regions.}
\label{tab:agn_fraction}
\begin{tabular}{c c c c c c c c c c c c}
\toprule
\multirow{2}{*}{$\lambda_*$} & \multirow{2}{*}{Redshift bin} & \multicolumn{5}{c}{Central region ($r < 0.2\,\mathrm{Mpc}$)} & \multicolumn{5}{c}{Outskirt region ($0.2 < r < 1.0\,\mathrm{Mpc}$)} \\
\cmidrule(lr){3-7} \cmidrule(lr){8-12}
 & & AGN fraction & Error & $N_{AGN}$ & $N_{gal}$ & $N_{group}$ & AGN fraction & Error & $N_{AGN}$ & $N_{gal}$ & $N_{group}$ \\
\midrule
> 10 & 0.0-1.0 & 0.024 & 0.006 & 20 & 821 & 115 & 0.014 & 0.002 & 67 & 4654 & 130 \\
 & 1.0-2.0 & 0.014 & 0.004 & 10 & 734 & 170 & 0.018 & 0.002 & 61 & 3309 & 191 \\
 & 2.0-3.0 & 0.013 & 0.007 & 4 & 301 & 93 & 0.010 & 0.003 & 14 & 1424 & 101 \\
 & 3.0-4.0 & 0.000 & 0.000 & 0 & 95 & 36 & 0.005 & 0.004 & 2 & 385 & 41 \\
\bottomrule
\end{tabular}
\end{table*}

There is a total of 15 HLAGN across the 9 richest galaxy groups sample.  The closest to the center AGN is in group ID~118 at $z=2.490$, with a distance of $\sim 72\,\mathrm{kpc}$. The furthest HLAGN is in group ID~82 at $z=2.9$, with a distance of $\sim 805\,\mathrm{kpc}$.

\begin{itemize}
    \item High Redshift ($z>1$):
    This category shows a wide range of locations. It contains the most centrally located AGN in the entire sample (Group~118 at $\sim 72\,\mathrm{kpc}$) as well as several AGN found in the outskirts (e.g., Groups~7, 121, and 82, with distances up to $\sim 805\,\mathrm{kpc}$).

    \item Low Redshift ($z<1$):
    Most AGN in this category are located in the outskirts of their host Groups. There are a few exceptions with relatively central locations, such as the AGN in Groups~1 and 4, at $\sim 131\,\mathrm{kpc}$ and $\sim 123\,\mathrm{kpc}$, respectively.
\end{itemize}
Out of the 15 HLAGNs identified, five are located in the central regions of their respective galaxy groups: 118 ($\sim 72\,\mathrm{kpc}$), 7 ($\sim 104\,\mathrm{kpc}$), 4 ($\sim 123$ and $\sim 177\,\mathrm{kpc}$; 2×), and 1 ($\sim 131\,\mathrm{kpc}$).
HLAGN populate both the central and outer regions of groups across all redshifts. The fraction peaks near $z \sim 2$, coincident with cosmic noon when abundant cold gas and frequent interactions funnel material toward supermassive black holes. The subsequent decline toward $z < 1$ implies dwindling fuel supplies and more effective feedback, although gas can still be retained in dense cores where close encounters and mergers recycle stripped material, sustaining residual activity.
\subsubsection{Large scale structure classification in rich groups}\label{sec:lss}
To quantify the environmental dependence of galaxy groups, we cross-matched the COSMOS-Web richest group catalog with the Taamoli et al. (in preparation) large-scale structure (LSS) classification, identifying systems residing in clusters, filaments, and the field. Taamoli measured the mean density contrast, account for the local background and for masked regions around bright sources, ensuring consistent sampling across environments. The resulting normalized densities were then compared across the three LSS categories to trace how the signal varies from the densest cluster cores to the more diffuse field regions (Taamoli et al., in preparation).

Cluster-like systems ($\texttt{LSS\_Flag}=2$) have a mean radius of $\sim592$\,kpc, filaments ($\texttt{LSS\_Flag}=1$) average $\sim475$\,kpc, and field-like associations ($\texttt{LSS\_Flag}=0$) extend to $\sim250$\,kpc, reinforcing the link between environment classification and physical scale.

\begin{figure}
    \centering
\includegraphics[width=0.45\textwidth]{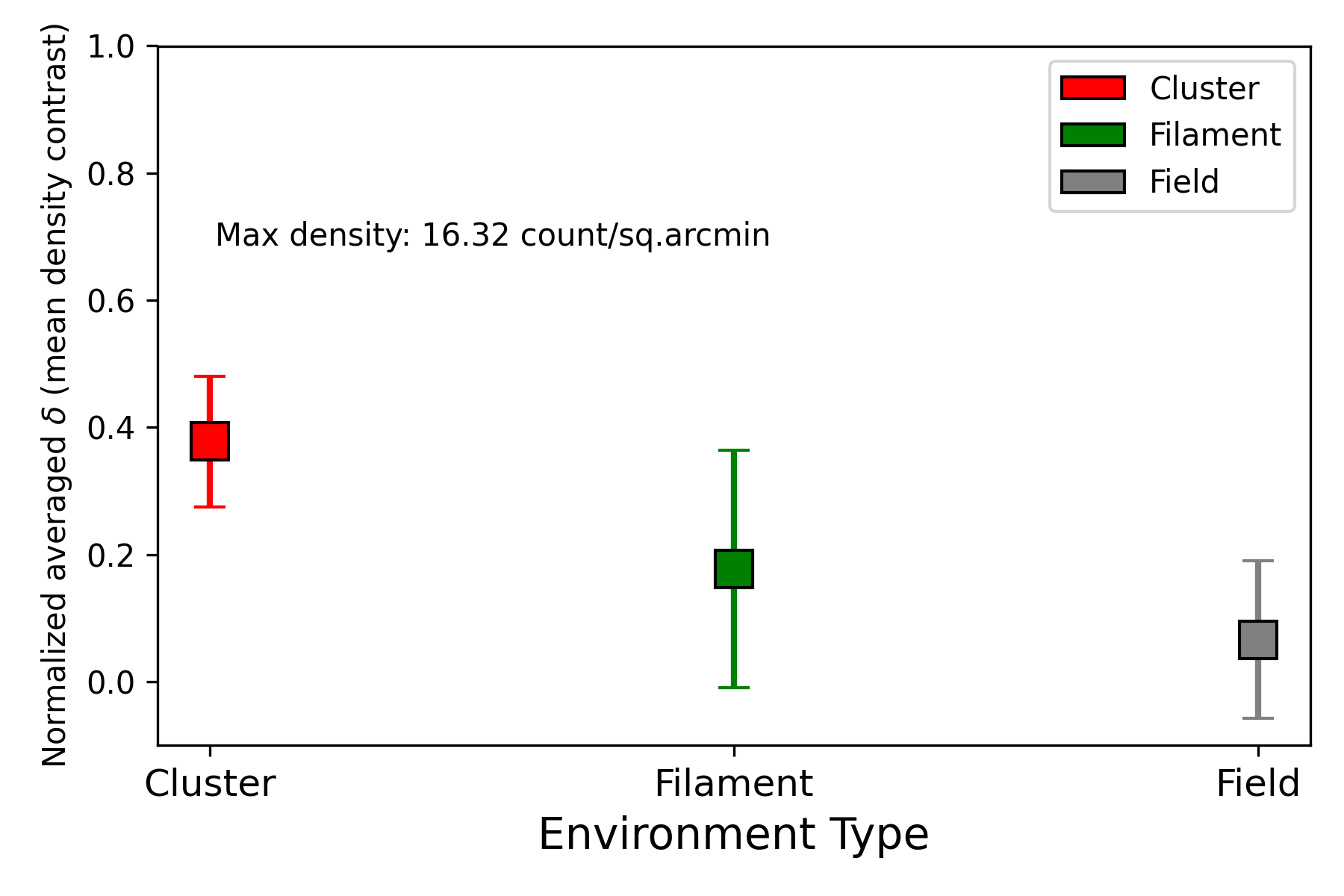}
    \caption{Normalized average surface density for groups as a function of environment classification. Squares with error bars indicate the mean and uncertainty for cluster, filament, and field systems, normalized to the peak density of group ID~4.}
\label{fig:density-env}
\end{figure}

We further stack groups within each category and normalize the profiles to the densest system in our sample, highlighting a clear gradient in both scale and density with environment.
 Groups in cluster-like regions exhibit the highest normalized mean densities, while filamentary systems bridge the transition toward the field population (Fig.~\ref{fig:density-env}). The average density decreases steadily from clusters to filaments and to the field, consistent with previous findings that link galaxy density to cosmic connectivity \citep{Castignani2022, Cautun2014}. Filaments represent environments of intermediate density between clusters and the field \citep{Cautun2014}, where in our case the larger uncertainty may be influenced by the limited sample size.


\subsection{The Red$-$Sequence population in rich groups}\label{sec:red-seq}
To predict the Red Sequence (RS) locus in the CMD across redshift range $0\,\lesssim\,z\,\lesssim\,4$, we construct a forward Stellar Population Synthesis (SPS) model using \texttt{Bayesian Analysis of Galaxies for Physical Inference and Parameter EStimation (Bagpipes)}\footnote{https://bagpipes.readthedocs.io/} \citep{Carnall2019}.
The model accounts for stellar mass, redshift, and environmental dependencies.
For each of the 25 groups considered in this analysis, we define the following physical quantities: a mass–metallicity relation (MZR; \citealt{Sanders2015, Sanders2021}), formation redshift ($z_{\rm form}$; is the redshift at which a galaxy formed 50\% of its stellar mass), star formation history (SFH), dust emission, and filter selection. We extend the \citet{Sanders2021} MZR to include redshift evolution and environmental metallicity boosts, while maintaining consistency with local observations \citep{Tremonti2004, Gallazzi2005}. The resulting relation is:

\begin{equation}
\begin{split}
12 + \log(\mathrm{O/H})_{\rm custom} = {} &
\left[\,8.69 + 0.64\,
\frac{\log M_\star - 10.4}
     {1 + \left(10^{\log M_\star - 10.4}\right)^{-1.2}}
+ \Delta_z\,\right] \\
& \times \mathrm{env\_boost}.
\end{split}
\end{equation}
where 8.69 is the $z=0$ field-galaxy asymptotic metallicity at the turnover mass, $M_{\rm TO} = 10.4$, with 0.64 representing the low-mass slope ($\gamma$); this $M_{\rm TO}$ is fixed slightly higher than the nominal value in \citet{Sanders2021} to better match massive-galaxy observations.
The exponent $-1.2$ reproduces the smooth curvature of the transition region in a computationally convenient way, effectively replacing the $\Delta$ parameter of the original formula. The redshift evolution term $\Delta_z$ and environmental boost $\mathrm{env\_boost}$ represent redshift evolution and environmental metallicity boosts, respectively.
Redshift evolution is calibrated against high-redshift measurements \citep{Maiolino2008, Cullen2014, Cullen2019} and recent JWST constraints \citep{Li2023, Curti2024, Stanton2025}, with the adopted linear form summarized in Appendix~\ref{sec:redshift_evolution_and_metallicity_boost}.
Redshift evolution is calibrated against high-redshift measurements and recent JWST constraints \citep{Maiolino2008, Cullen2014, Cullen2019}, with the adopted linear form summarized in Appendix~\ref{sec:redshift_evolution_and_metallicity_boost}. This term captures the systematic decrease of metallicity with increasing redshift, reflecting both high-redshift galaxy observations and theoretical expectations for chemical enrichment across cosmic time.
This formula preserves the low-mass slope, turnover mass, and curvature of the original Sanders MZR while allowing physically motivated extensions. Observations indicate that galaxies in groups and clusters are more metal-rich than field galaxies at fixed stellar mass \citep{Pasquali2012, Peng2015}. We implement this as a multiplicative boost to the metallicity, which depends on stellar mass and decreases with redshift, reflecting the reduced efficiency of environmental processing in the early Universe \citep{Fossati2017, Chartab2020}. Field galaxies are kept as the baseline, with no enhancement. The environmental metallicity boost applied to group galaxies is described in Appendix~\ref{sec:redshift_evolution_and_metallicity_boost}. This factor is multiplicative on the base mass–metallicity relation \citep{Fossati2017, Chartab2020}.
We incorporate the fundamental metallicity relation \citep{Mannucci2010}, ensuring consistency between stellar mass, metallicity, and star-formation rate.
A primordial metallicity floor of $\log(\mathrm{O/H}) = 7.5$ is imposed, and abundances are expressed relative to the solar scale of \citet{Asplund2009} ($12 + \log(\mathrm{O/H})_{\odot} = 8.69$).
The galaxy formation redshift $z_{\rm form}$ is assigned based on stellar mass and observation redshift following downsizing trends \citep{Thomas2010, McDermid2015, Behroozi2013, Rodriguez-Gomez2016, Nakajima2023, Curti2024} (see Appendix~\ref{sec:formation_redshift} for the parameterization). The assigned $z_{\rm form}$ is limited to $z_{\rm obs} + 2$ and capped at 10; formation times shorter than $0.1\,\mathrm{Gyr}$ are avoided.

Dust attenuation for red sequence galaxies is modeled as a function of stellar mass and redshift, following observational constraints \citep{Whitaker2017, Bouwens2016, Fudamoto2021}. More massive galaxies exhibit slightly higher V-band attenuation, while attenuation increases moderately with redshift. To account for environmental effects, a small reduction is applied for galaxies in dense regions, reflecting processes such as gas stripping (see Appendix~\ref{sec:dust} for the parameterization). The resulting prescription ensures low but evolving dust levels, with an upper limit of $A_V = 0.3\,\mathrm{mag}$, consistent with the typically quiescent nature of red sequence systems.

The age spread of red sequence galaxies is modeled through an exponentially declining SFH characterized by a timescale $\tau$, which depends on stellar mass and redshift \citep{Thomas2010, McDermid2015}. More massive galaxies are assigned shorter $\tau$, reflecting their rapid, early formation, while lower-mass systems have more extended SFHs. The timescale also evolves mildly with redshift, allowing slightly longer star formation at earlier epochs (see Appendix~\ref{sec:sfh_timescale} for the parameterization).
A minimum value of $\tau = 0.05\,\mathrm{Gyr}$ is imposed to ensure physically realistic evolution. This prescription captures the observed downsizing trend and provides realistic age spreads for synthetic red sequence galaxies.
For each group, we generate a realistic red sequence population by sampling galaxies across a wide stellar mass range, $\log_{10}(M_\star/\modot)=7.5$–$12.0$. This range captures both low-mass satellites and the most massive central galaxies, allowing a physically motivated modeling of the red sequence.
To predict observable colors, we select filter pairs that bracket the redshifted 4000\,\AA\ break, adapting to the cluster redshift. For low-redshift clusters ($z_{\rm cl}<0.7$), we use HSC $g$ and $z$ bands. Intermediate redshifts ($0.7\leq z_{\rm cl}<1.4$) employ HST F814W and JWST F277W filters, while higher redshifts ($1.4\leq z_{\rm cl}<4$) adopt JWST F115W and F277W. This ensures that the chosen filters effectively trace the spectral feature that defines quiescent galaxies at each epoch.

The predicted location of the red sequence on the CMD for group ID~4 is shown in Fig.~\ref{fig:combined_plots_sfr_ccd_cmd} (right panel) and for a subsample is shown in the Appendix (Fig.~\ref{fig:combined_plot_sfr}, while the remaining plots are available online via the Zenodo link: \url{https://zenodo.org/uploads/17407954}).
The orange line and pink shaded area on the CMD diagram show the predicted red-sequence location at the group's redshift $z_{\mathrm{DET}}$, with
$\pm0.25\,\mathrm{mag}$ reflecting the uncertainty in the color. We identify clear red sequences (RS) in 5 out of 25 galaxy groups, prominently established at $z < 1$, with early indications of RS features in 3 additional groups up to $z \sim 2.2$.
 The groups with the significant ETGs and LTGs in the place of red sequence, are IDs~1(27\,/\,42), 35(17\,/\,23), 36(15\,/\,4), 4(8\,/\,3), 15(9\,/\,41), 304(6\,/\,5), 85(3\,/\,4), 20(4\,/\,2).
For group ID~4, we found 11 out of 84 galaxies ($\sim13\%$) on the red sequence, where the percentage is relative to all group members.
For all 25 groups we found an average fraction of $\sim\,10\%$ (105/1075) ETGs and $\sim\,13.5\%$ (145) LTGs on the red sequence. We found that 19 out of 25 groups contain fewer than five ETG galaxies in the red sequence region. Significant presence of LTGs ($\sim$\,13.5\%) on the red sequence is indicative of galaxies that have quenched their star formation while retaining their disk-dominated morphology. This observation is consistent with studies that track the morphological evolution of quiescent galaxies, such as \citet{Cerulo2017}, who, using the HAWK-I Cluster Survey (HCS) sample at $0.8<z<1.5$, document a $\sim$\,20\% increase in the fraction of S0 galaxies on the red sequence from $z \sim 1.5$ to $z \sim 0.05$.
These results suggest that the LTGs observed in our high-redshift groups are the progenitors actively undergoing morphological transformation into S0 galaxies as they evolve toward the present day \citep{Cui2024, alda2025}.
Using Hyper Suprime-Cam Subaru Strategic Program
(HSC-SSP) data at $0.3<z<0.6$, \citet{Shimakawa2021} found a deficit of spiral galaxies in dense environments and clusters, consistent with the emergence of the morphology–density relation. In contrast, our COSMOS-Web groups still host a high fraction of LTGs in dense regions, suggesting we are observing an earlier stage where quenching is underway but morphological transformation has not yet taken hold.
We further illustrate the locations of the ETG and LTG populations in the Color--Color Diagram (CCD) (Fig.~\ref{fig:combined_plots_sfr_ccd_cmd}, middle panel) using empirically defined trapezoidal segments. The combination of HST optical (F814W) and JWST NIR filters (F277W, F444W) provides sufficiently broad wavelength coverage to trace the full range of galaxy colors across our redshift interval.
The part that clearly depends on redshift is the horizontal line before the ankle point. This line becomes redder at higher redshift: it starts at color $\approx 0.5$ at lower $z$, increases to $\approx 1$--$2$ around $z \sim 1$, and reaches $\approx 2.5$ at higher redshifts, reflecting the gradual aging of stellar populations and the intrinsic evolution of galaxy colors over cosmic time.
In contrast to fixed rest-frame color cuts \citep[e.g.,][Fig.~3]{Ilbert2013} or redshift-dependent prescriptions \citep[e.g.,][Fig.~4]{Ghaffari2021}, this CCD-based method isolates the region where morphological transformation lags behind color quenching, allowing us to identify LTGs that maintain disk-like morphology while already exhibiting redder colors.
The shaded background regions define two principal color-selected populations, each marked with corresponding red and green box outlines across all three panels. The red-shaded zone in the upper left corresponds to the quiescent population, which includes both ETGs and LTGs. The green-shaded region, located between the horizontal cut and the inclined boundary, delineates the Red Star-forming (RSF) population, which is primarily composed of LTGs.
These same populations are highlighted in the log($\text{SFR}$)--log($\text{M}_{\star}/\modot$) diagram (Fig.~\ref{fig:combined_plots_sfr_ccd_cmd}, left panel) using the same color and symbol coding as in the middle panel. Galaxies within the red box occupy the region below the Main Sequence (MS), approaching the fully quenched regime. In contrast, galaxies within the green box lie closer to, or just below, the MS, indicating that they remain actively star-forming despite having colors redder than those of the typical blue, star-forming population.
 Additional examples are shown in Fig.~\ref{fig:combined_plot_sfr} and available online (\url{https://zenodo.org/uploads/17407954}). Whether the observed redness is due to dust obscuration or to the onset of quenching is beyond the scope of the present paper.
 We compute the evolution of RS galaxies by dividing the number of galaxies that lie above the red sequence minus 0.25 magnitudes (RS$-0.25$) in the CMD by the total number of filtered galaxies within the corresponding redshift bin. We restrict the spatial distance to $500\,\mathrm{kpc}$ to exclude members with probability $< 0.5$, while results within $800\,\mathrm{kpc}$ show a similar general behavior.
In Fig.~\ref{fig:quenched_fractions_rs}, we show the mean RS fractions as a function of redshift. The mean fraction decreases steadily from $z \sim 0.5$ to $z \sim 3.5$. At $z \sim 0.5$, the ALL population reaches about $40\%$, with LTGs contributing a higher fraction (about $25\%$) than ETGs (about $17\%$).
By $z \sim 1.5$, the ALL fraction falls below $15\%$, and both ETG and LTG fractions continue to decline, converging to very low values (3–8\%) by $z \sim 3.5$.

\begin{figure*}[htbp]
    \centering
    \begin{subfigure}[b]{0.32\textwidth}
        \centering
        \includegraphics[width=\textwidth]{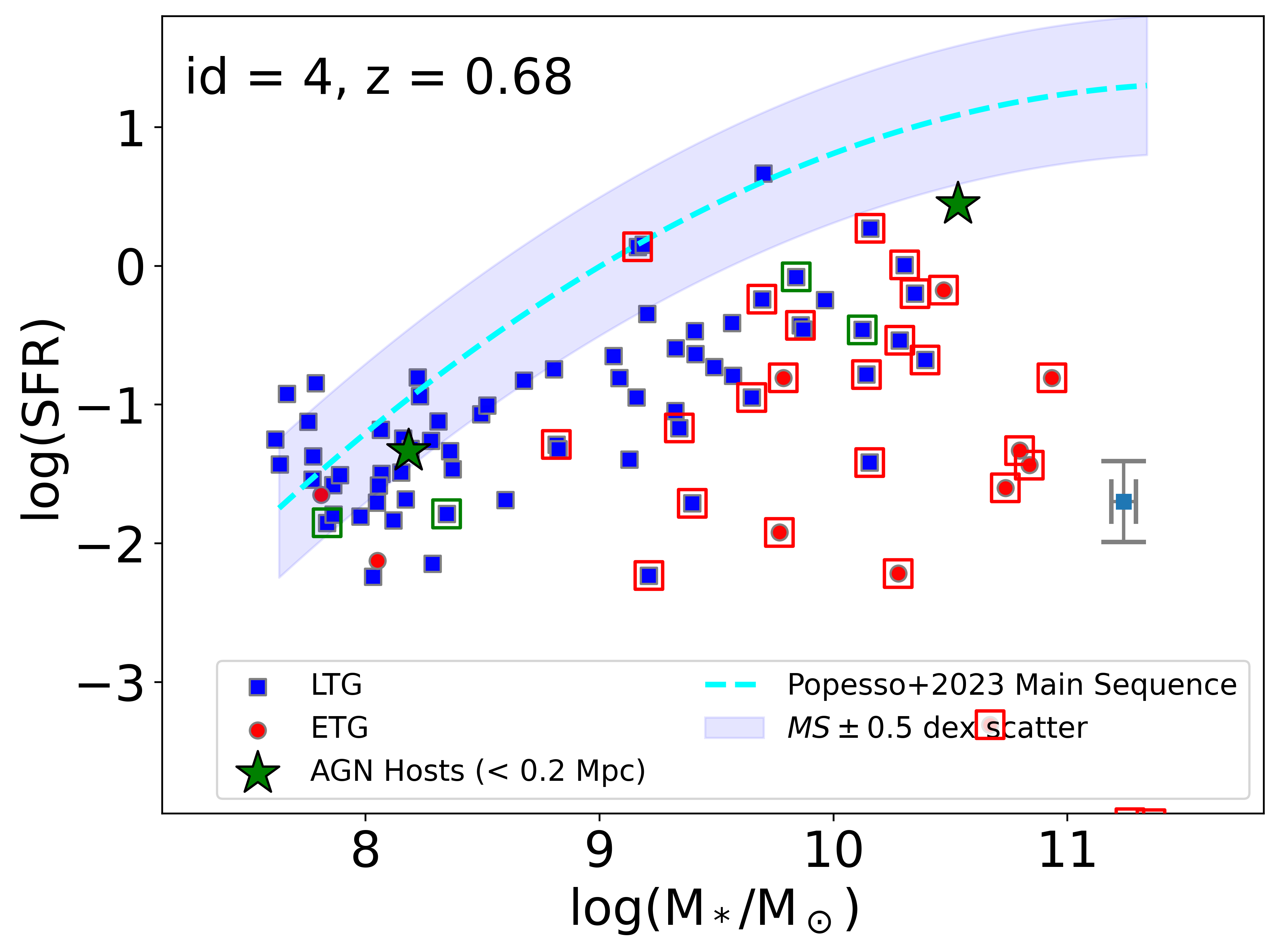}
    \end{subfigure}
    \hfill
    \begin{subfigure}[b]{0.32\textwidth}
        \centering
        \includegraphics[width=\textwidth]{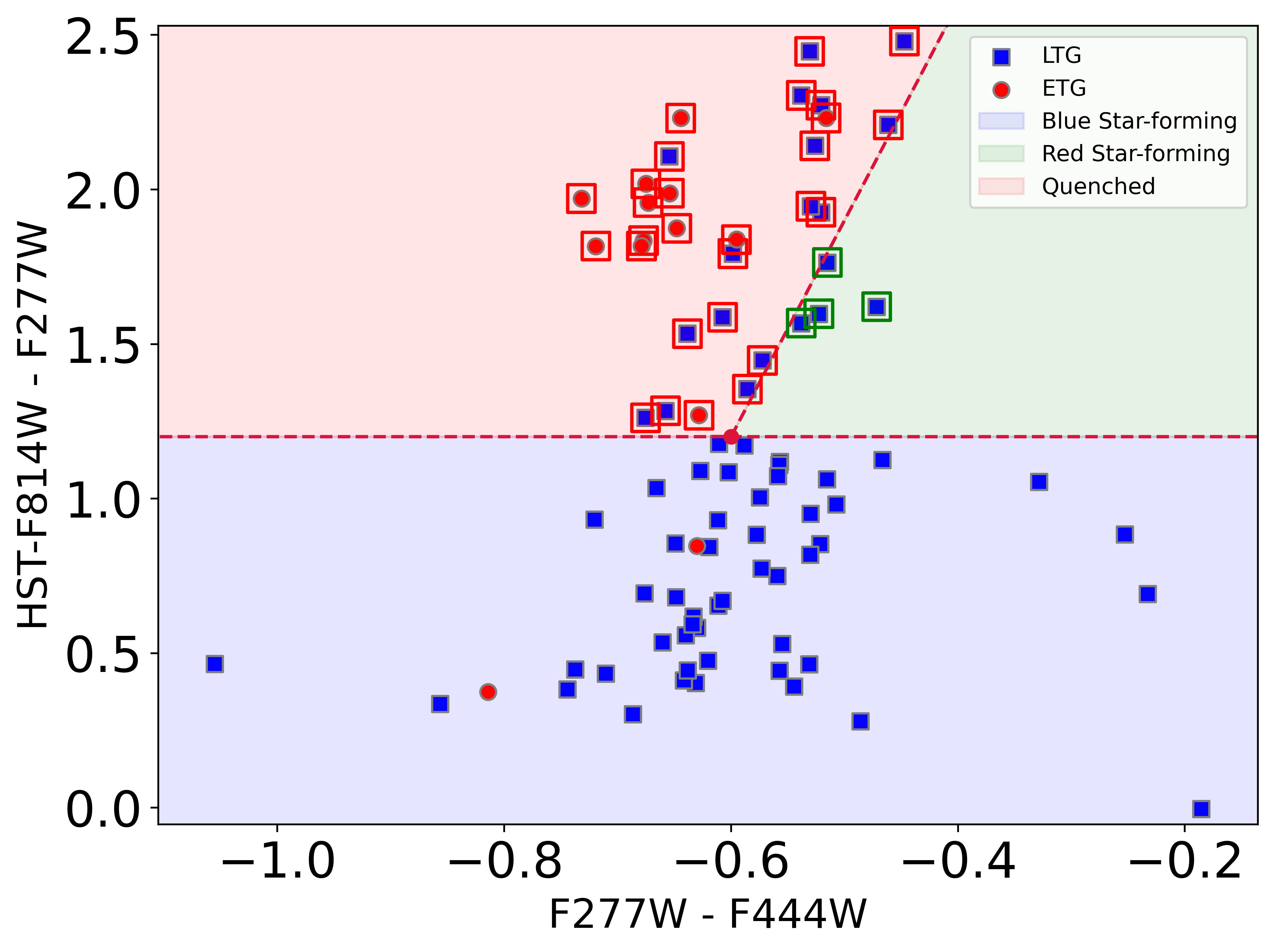}
    \end{subfigure}
    \hfill
    \begin{subfigure}[b]{0.32\textwidth}
        \centering
        \includegraphics[width=\textwidth]{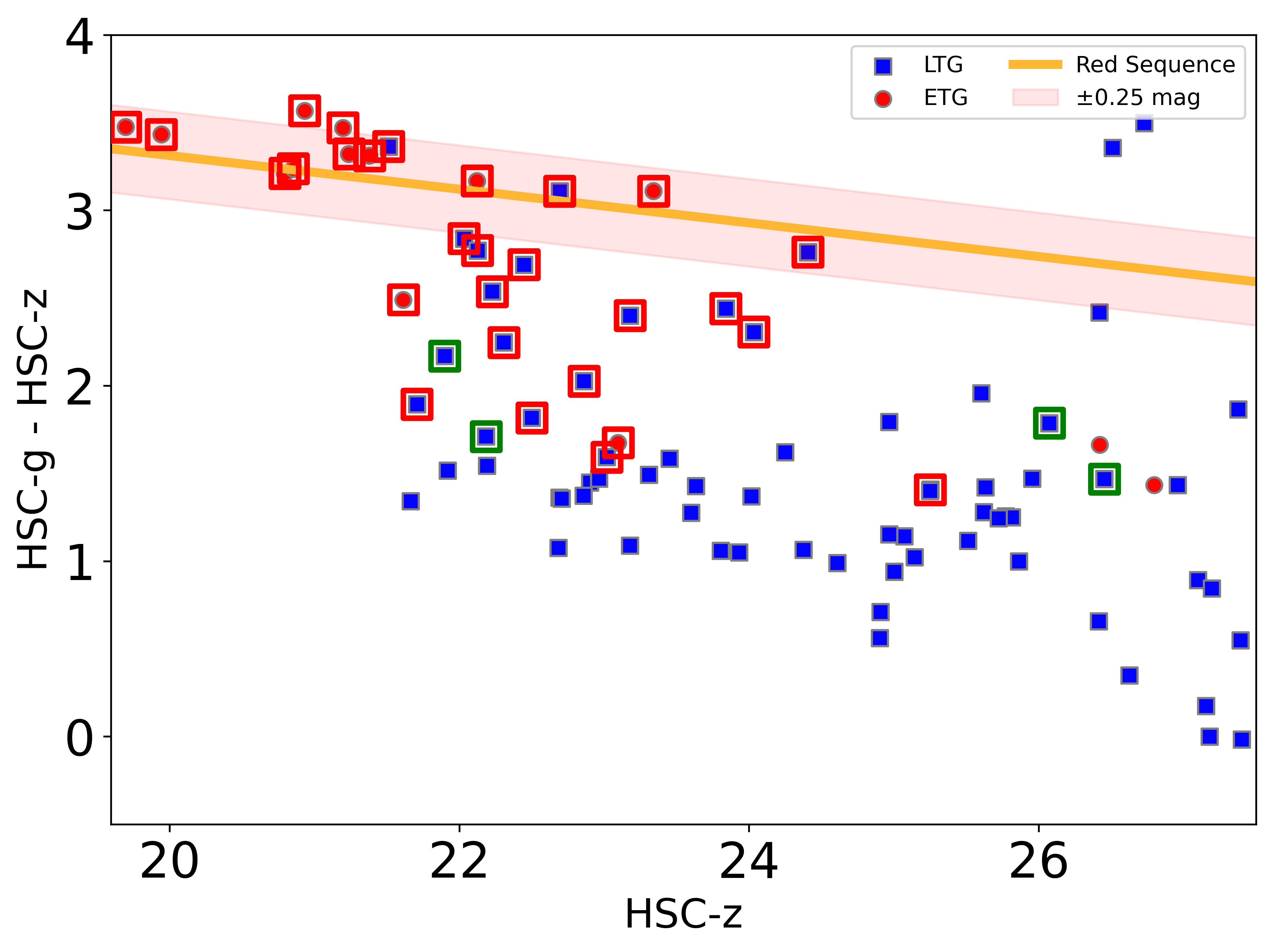}
    \end{subfigure}
    \caption{Diagnostic diagrams for group ID~4. Left: SFR versus stellar mass with LTGs (blue squares) and ETGs (red circles); the gray cross indicates typical uncertainties and the cyan dashed line shows the \citet{Popesso2023} main sequence. Middle: rest-frame color–color diagram. A crimson dashed line segments define an empirical, trapezoidal boundary separating the galaxy population into Blue Star-forming, Red Star-forming (green-shaded), and Quiescent (red-shaded) regions. To highlight the key non-blue populations, galaxies are marked on all panels with a thick green square outline (Red Star-forming) or a thick red square outline (Quiescent). Right: color–magnitude diagram with orange line and pink shaded area marking the predicted red-sequence locus at $z_{\mathrm{DET}}\pm0.25\,\mathrm{mag}$.
    }

    \label{fig:combined_plots_sfr_ccd_cmd}

\end{figure*}
\subsection{Stellar mass and star
formation in rich groups}\label{sec:quenching}
Overall, the emerging picture at high redshift indicates two complementary modes of quenching:
massive galaxies experience rapid, internally driven quenching, while lower-mass galaxies undergo delayed, environment-driven quenching as clusters assemble and virialize from $z \sim 2$ to $z \sim 1$.
We investigate the evolution of the ETGs/LTGs in Starburst, main sequence (MS), and Quiescent regions across the widest redshift range ($0<z<3.6$) and stellar mass interval ($10^{7.5}$–$10^{11.5}\,\modot$) probed to date (Fig.~\ref{fig:mass_completeness}). We apply a stellar-mass threshold of $\log(M_{\text{cut}}(z))$ to define the final sample, ranges from $6.5$ at $z=0$ to $8.9$ at $z=4.0$.

\subsubsection{The $M_{\star} - \mathrm{SFR}$ relation}
The star-forming main sequence (SFMS) links galaxy stellar mass and SFR, providing a baseline to identify typical versus unusual star-forming activity. Its slope and normalization evolve smoothly with redshift, reflecting mass-dependent growth rather than stochastic events, making it a key tool to study galaxy evolution \citep{Speagle2014}.
\citet{Popesso2023} compiles a comprehensive census of 64 star-forming galaxies ($0<z<6$, $10^{8.5}$–$10^{11.5},\modot$) and showed that the SFMS is best described by a two-parameter function: a redshift-dependent normalization and a turnover mass $M_{0}(t)$. Below $M_{0}(t)$, galaxies maintain constant specific star formation rate (sSFR), while above it the sSFR is suppressed, producing the characteristic SFMS bending. This framework links $M_{0}(t)$ to the transition in halo accretion modes and provides a formalism to locate ETGs and LTGs relative to the evolving SFMS.
\begin{equation}
\begin{split}
\log_{10}(\text{SFR}) = & \left[ (0.84 \pm 0.02) - (0.026 \pm 0.003)t \right] \log_{10}(M_\star) \\
& - \left[ (6.51 \pm 0.24) - (0.11 \pm 0.03)t \right] \pm 0.23
\end{split}
\end{equation}
Where $t$ is the cosmic time. This is important to understand how these fundamental galaxy properties are influenced by the group environment as they evolve across cosmic time.
To reliably separate star-forming regions in the $M_\star$–SFR plane (see left panel of Fig.~\ref{fig:combined_plots_sfr_ccd_cmd}), we define the main sequence (MS) as the locus within $\pm 0.5$\,dex of the SFMS at the group's redshift a bit wider than what is accepted typically for clusters (e.g., $\pm 0.3$\,dex, \citet{Janowiecki2020, Popesso2023, Finn2023, DeDaniloff2025}). We define the starburst region as lying above the MS and the quiescent region as lying below it.
In order to distinguish the evolution of ETGs and LTGs in different sections of SFMS, we stacked all the group members to achieve a statistically robust sample. The galaxies were then sorted into a grid of bins based on their stellar mass (High-Intermediate-Low) and redshift ($\Delta z = 0.5$). For each bin, we calculate the fraction of ETGs and LTGs for three distinct populations: the MS, Starburst (SB), and Quiescent (Q).
\subsection{Quenching and morphological evolution across mass bins}\label{sec:quenching_mass_bins}

\begin{figure}[t!]
    \centering
    \includegraphics[width=0.9\linewidth]{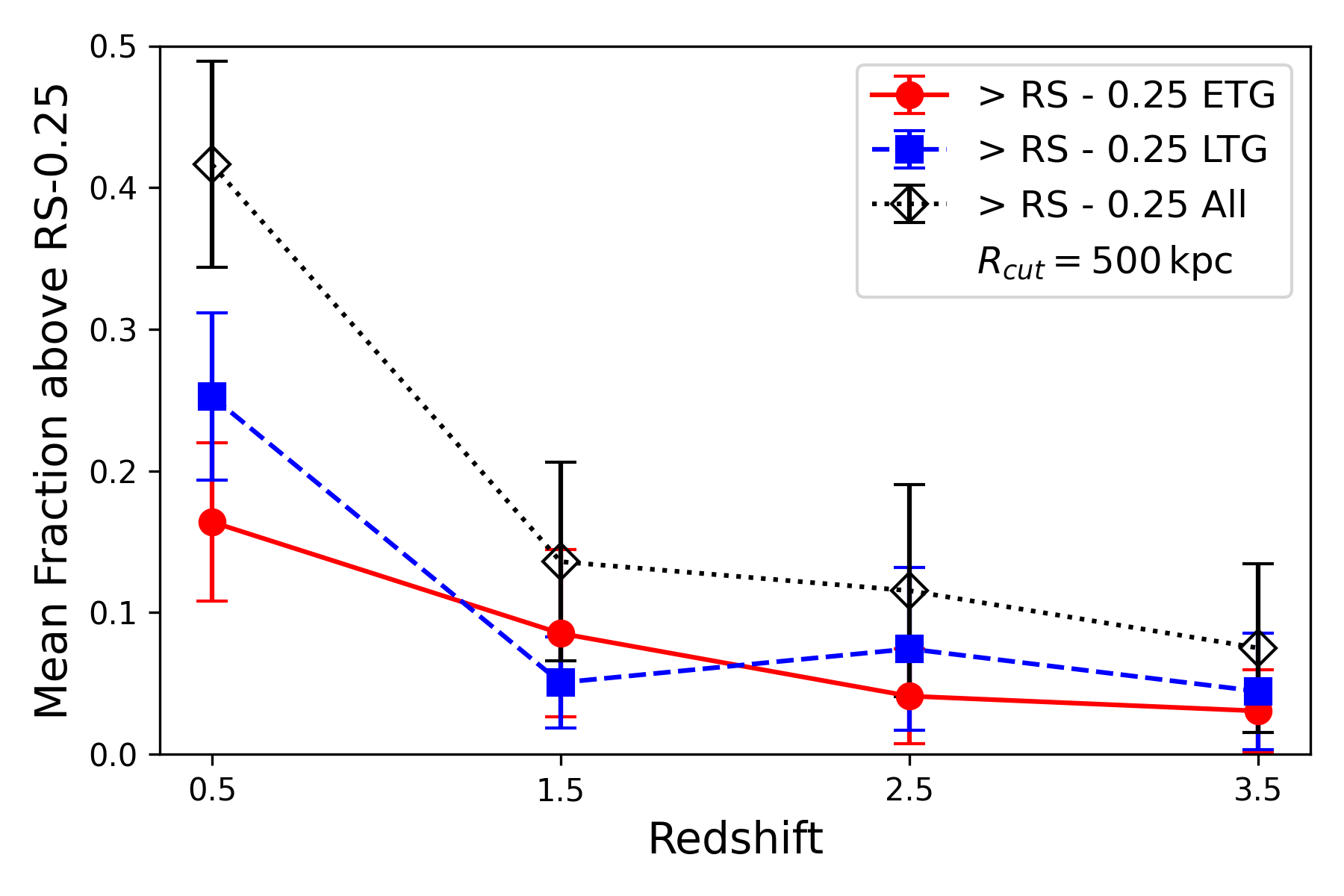}
    \caption{Mean fraction of ALL, ETG, and LTG galaxies as a function of redshift, shown for galaxies selected above the RS minus $0.25~\mathrm{mag}$ in the CMDs.}
    \label{fig:quenched_fractions_rs}
\end{figure}

\begin{figure}[ht!]
\centering

    \includegraphics[width=0.24\textwidth, height=0.20\textwidth]{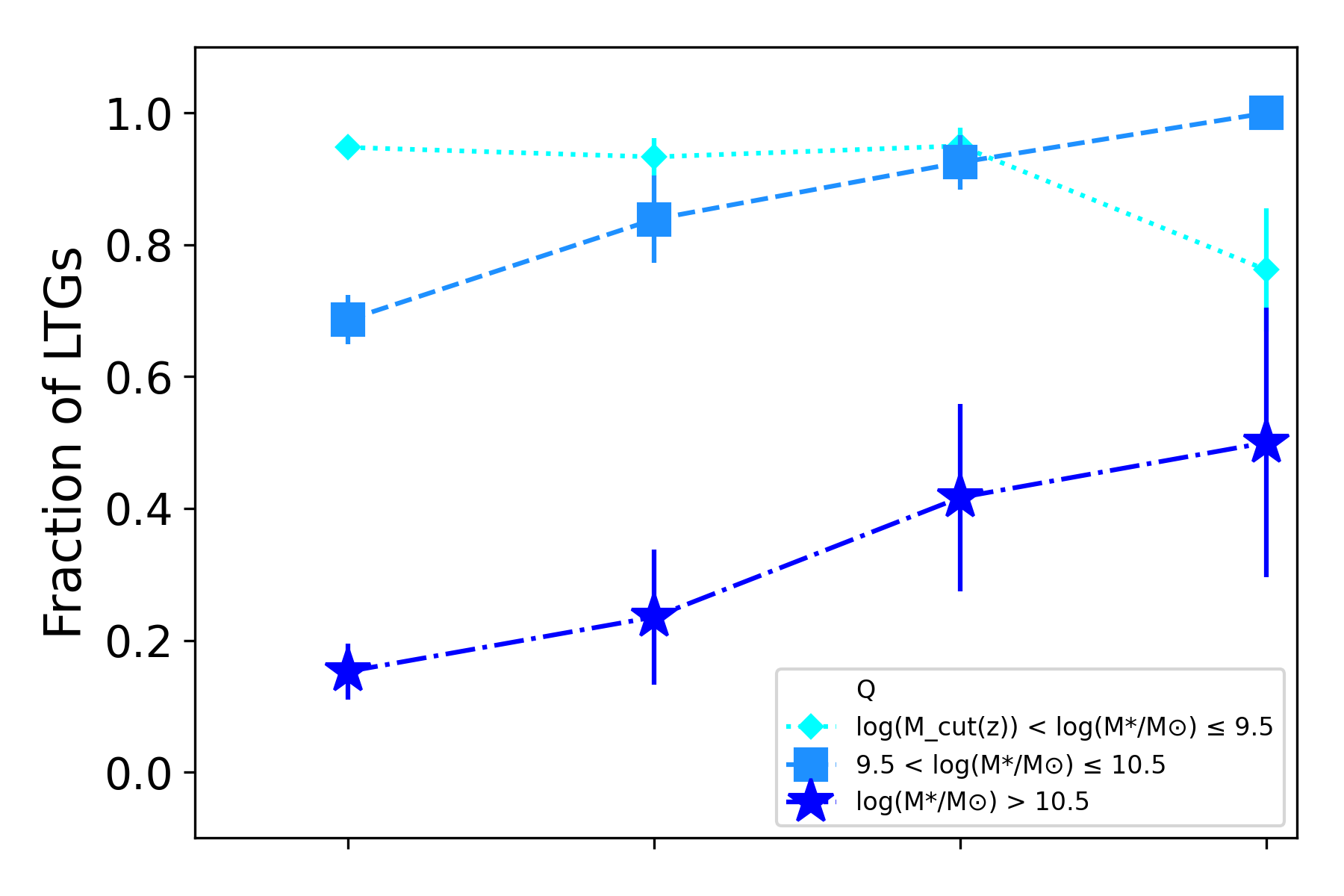}
    \includegraphics[width=0.23\textwidth, height=0.20\textwidth]{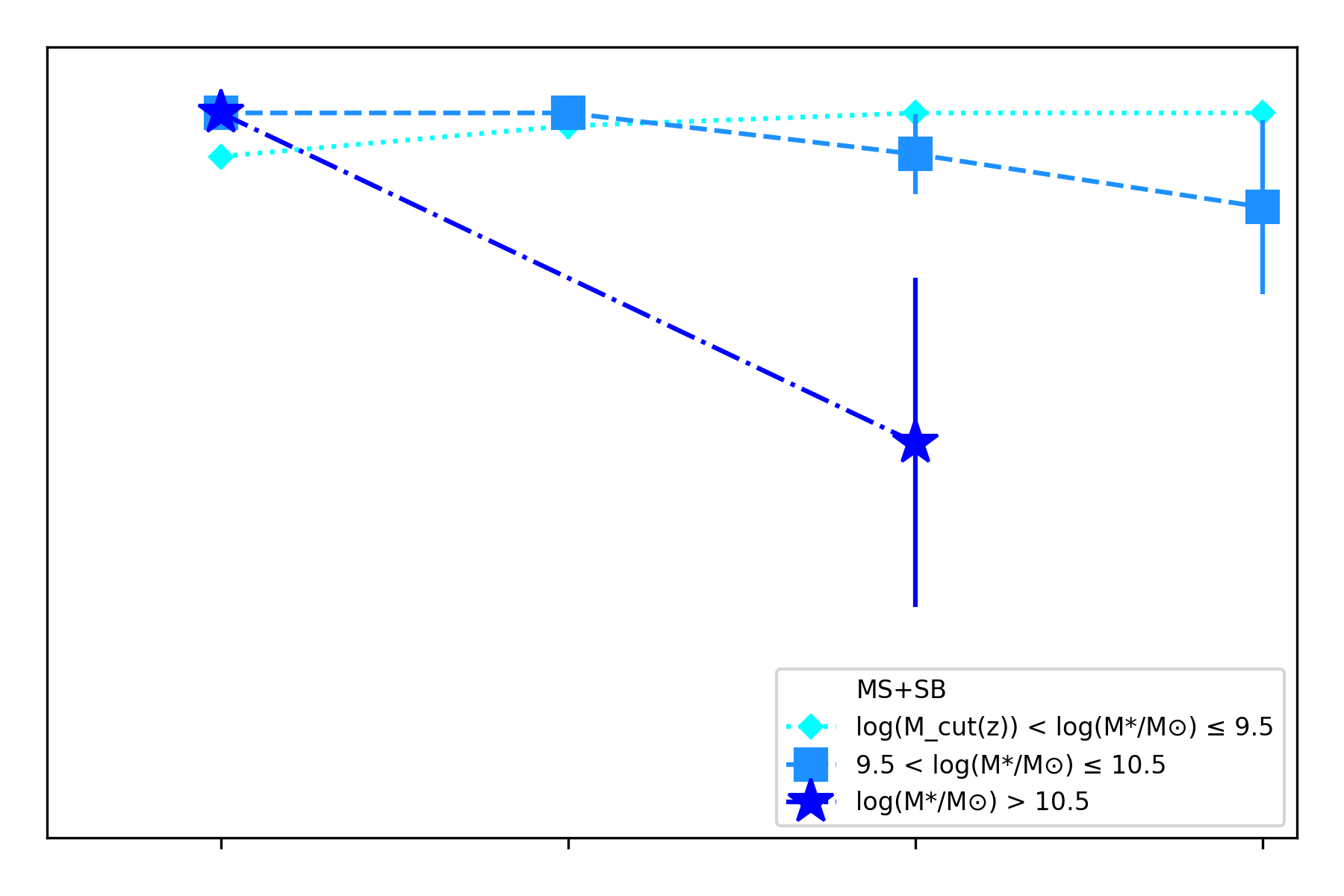}

    \includegraphics[width=0.24\textwidth, height=0.22\textwidth]{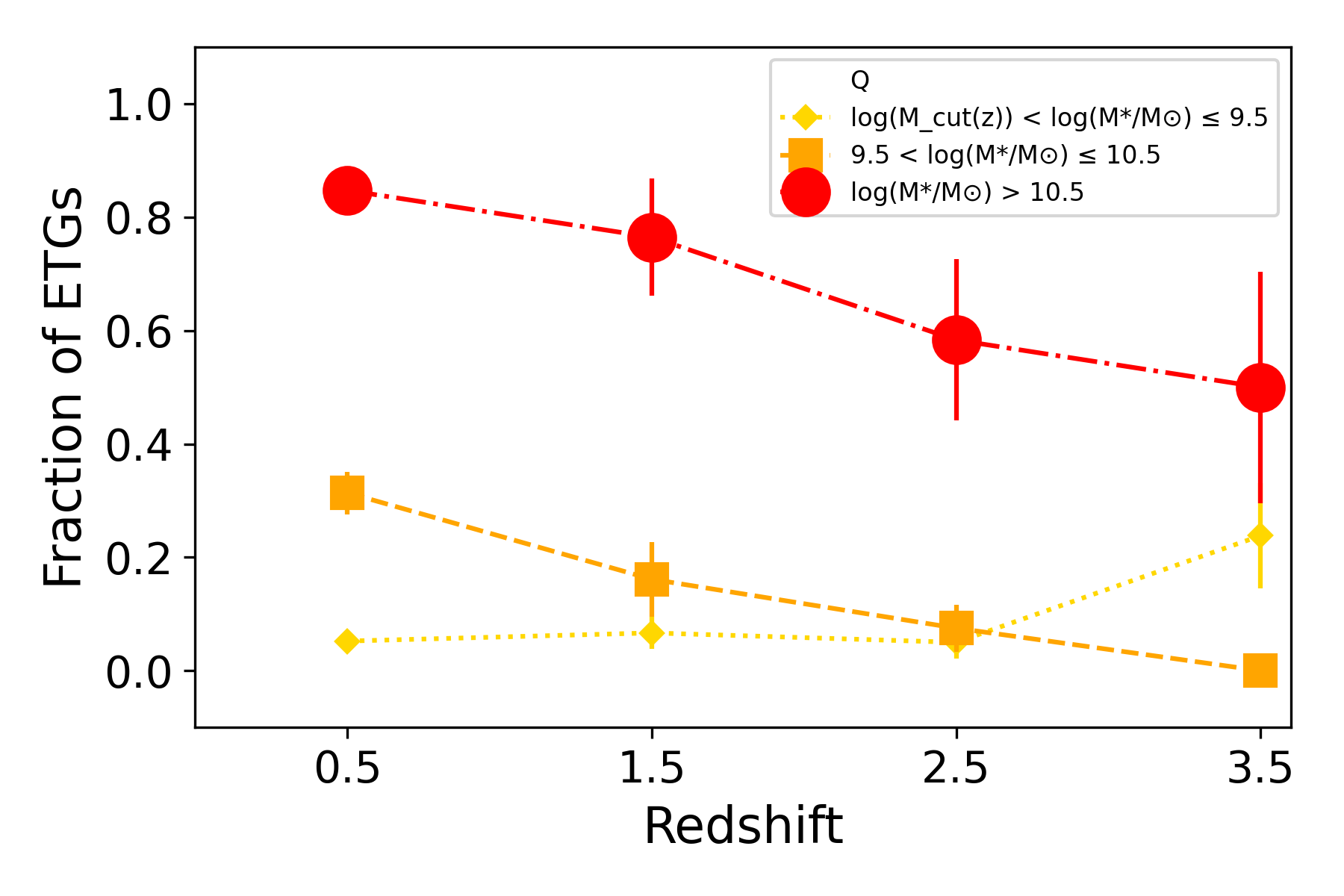}
    \includegraphics[width=0.23\textwidth, height=0.22\textwidth]{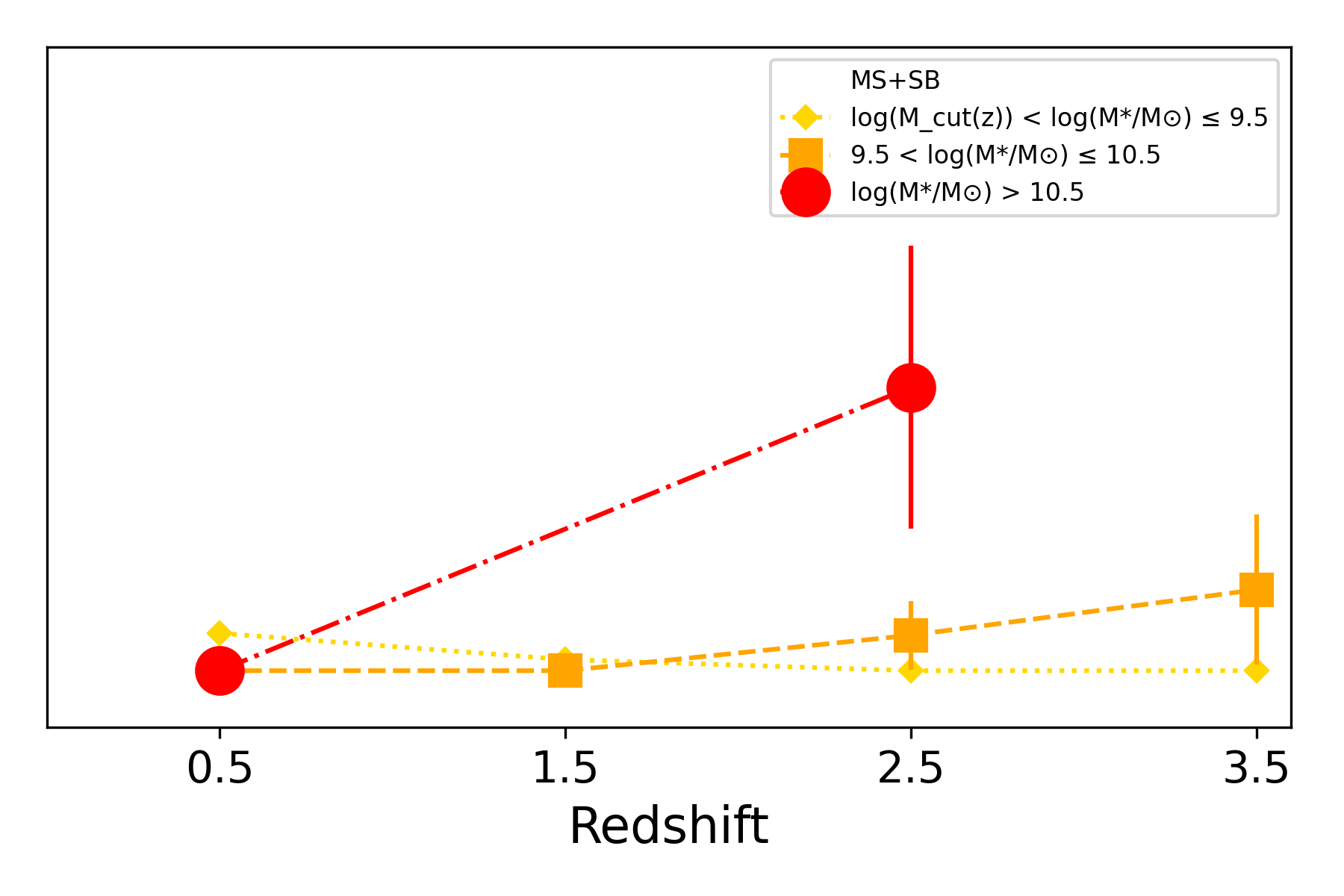}

 \caption{Evolution of LTG (top row) and ETG (bottom row) fractions within the Quiescent (Q), main-sequence$+$starburst (MS+SB) regimes (from left to right). Marker size encodes stellar-mass bins: $[\log(M_\star/\modot) < 9.5]$, $[9.5 \le \log(M_\star/\modot) \le 10.5]$, and $[\log(M_\star/\modot) > 10.5]$, and dashed error bars represent binomial uncertainties. We apply a stellar-mass threshold of $\log(M_{\text{cut}}(z))$ to define the final sample, ranges from $6.5$ at $z=0$ to $8.9$ at $z=4.0$.
}
    \label{fig:ms_evolution}
\end{figure}

Figure \ref{fig:ms_evolution} reveals the critical interplay between stellar mass, galaxy
morphology (ETG/LTG), and star-forming activity
across cosmic time.
In the High-Mass bin ($\log_{10} (M_\star/M_\odot) > 10.5$), quiescent galaxies transition
from being LTG-dominated ($\sim 50\%$) at high redshift ($z \approx 3.5$)
to ETG-dominated ($\sim\,80\%$) at low redshift ($z \approx\,0.5$). The Intermediate-Mass bin ($9.5 < \log_{10}(M_\star/\modot) < 10.5$) shows only a mild evolution:
the LTG fraction drops slightly from $\sim
\,90\%$ to $\sim
\,75\%$ ($z = 3.5$ to 0.5), corresponding to an increase in the ETG fraction from $\sim 0\%$ to $\sim 30\,\%$. The Low-Mass bin ($\log(M_{\text{cut}}(z)) < \log_{10}(M_\star/\modot) < 9.5$), is consistently dominated by LTGs, maintaining a fraction of $\sim 90\%$ across all redshifts. The corresponding ETG fraction remains low, staying within the range of $\sim 10\%$ to $\sim 20\%$.

The MS+SB LTG fraction is nearly universal (above $80\%$) for the low and intermediate mass bin across all redshifts, leaving the MS+SB ETG fraction negligible ($<20\%$). However, the High-Mass MS$+$SB bin presents a unique evolution: the LTG fraction increases significantly from $\sim60\%$ at $z>2$ to $\sim100\%$ at $z<1$. Correspondingly, the ETG fraction decreases sharply from $\sim40\%$ at $z>2$ to near $0\%$ at $z<1$.
Based on our findings, we propose the following quenching scenario:
\begin{itemize}
    \item High-mass galaxies: Rapid, transformative quenching, largely mass-driven.
    \item Intermediate-mass galaxies: Mild quenching, partially environment-driven.
    \item Low-mass galaxies: Mostly star-forming, environmental quenching is slow and inefficient.
\end{itemize}
Our results align with \citet{Peng2012} regarding mass quenching in high-mass galaxies and environmental effects on intermediate-mass satellites. However, for low-mass galaxies, we find evidence for slow quenching, whereas \citet{Peng2012} and recent local studies report higher quenching efficiencies. Specifically, \citet{Ann2025} show that in the local Universe ($z < 0.01$), dwarf galaxies in group and cluster environments undergo early quenching, while late quenching is prevalent in the field. This discrepancy may indicate that the environmental processing affecting local samples differs from that operating at higher redshift. High-mass galaxies experience rapid, transformative quenching, in agreement with findings from the GOGREEN clusters \citep{McNab2021, Gully2025}. Intermediate-mass galaxies exhibit mild, environment-driven quenching, consistent with a small excess of transition galaxies in clusters, while low-mass galaxies remain largely star-forming, indicating slower, less efficient environmental quenching than reported in GOGREEN. This may be due to GOGREEN probing cluster environments instead of groups. The observed evolutionary trend is broadly consistent with the higher quenching probabilities in dense environments reported by \citet{Taylor2023}.
\citet{DeDaniloff2025} report that in the Cl0024 cluster at $z\sim0.4$, $\sim35\%$ of star-forming galaxies are in a ``suppressed'' state, while field galaxies are mostly active. In contrast, our COSMOS-Web groups at $z\sim0.5$ show a strong stellar-mass dependence: $\sim80\%$ of massive galaxies ($M_\star>10^{10.5}\,\modot$) are quiescent, $\sim20\%$ at intermediate masses, and $<10\%$ at low masses ($\log(M_{\text{cut}}(z)) < \log_{10}(M_\star/\modot) < 9.5$). Unlike Cl0024, our classification includes both morphologically identified ETGs and LTGs, capturing fully quenched and suppressed populations, highlighting a faster quenching of massive satellites and ongoing star formation in low-mass systems.

In a recent \citet{Gentile2025} paper, the relationship between local density and quenching/morphological transformation was investigated up to $z \approx 1$ over an area exceeding 60~deg$^2$. The study found consistent results, namely that quenching precedes morphological transformation in dense environments. Furthermore, \citet{Hatamnia2025} studied the evolution of galaxies in LSS in the COSMOS-Web field up to $z \sim 7$ and found that stellar mass positively correlates with density up to $z \sim 5.5$, with mass-driven quenching dominant at $z \gtrsim 2.5$ and environmental quenching becoming increasingly effective for low-mass galaxies toward lower redshift.
Quenching follows two fundamentally different pathways, closely tied to stellar mass. In massive galaxies, we suggest that the rapid cessation of star formation occurs early and violently, transforming disks into spheroidal ETGs. This process may be partially driven by AGN feedback, along with other mechanisms such as violent disk instabilities or major mergers (see Sect.~\ref{sec:RSD} for a detailed analysis of the AGN fraction). At intermediate masses, quenching appears to proceed more slowly and is dominated by environmental processes, such as gas starvation or stripping, which suppress star formation while largely preserving disk structures, leaving behind passive LTGs. At the low-mass end, quenching remains highly inefficient: even when external mechanisms act, galaxies generally retain their disk morphology. Thus, we suggest that while massive galaxies follow a fast, transformative quenching route, lower-mass systems experience a slower, non-destructive pathway, illustrating the dual nature of galaxy evolution across cosmic time.

\section{Summary and Conclusion}\label{sec:summary}
Using the COSMOS-Web JWST data set we derived a motivated radius $R_\mathrm{cut}$ for each group by intersecting radial surface-density profiles with the local field background, finding that denser systems occupy larger radii than filamentary or field structures (see Sect.~\ref{sec:lss}) and that LTGs outnumber ETGs by roughly 2:1 within $100\,\mathrm{kpc}$; the wide COD range ($4$–$70$\,counts\,arcmin$^{-2}$) signals that many cores are still assembling or retain significant substructure.

Fifteen HLAGN are matched to the 25 richest groups, with four sources lying in the central $0.2\,\mathrm{Mpc}$ ($72–131\,\mathrm{kpc}$) and eleven inhabiting the outskirts ($177–834\,\mathrm{kpc}$); their fractional contribution peaks at $z\sim2$ for both groups and field galaxies, tracing the availability of cold gas at cosmic noon and indicating that interactions in group outskirts increasingly sustain black-hole growth at high redshift.

Central HLAGN hosts span starburst, main-sequence, and quiescent classes (1/2/2 respectively), demonstrating that residual gas reservoirs can continue feeding black holes after star formation quenches; individual cases such as the starburst AGN in Group~7 ($z=1.56$) and in Group~304 ($z=0.34$) illustrate how nuclear activity and intense star formation coexist across environments (Sect.~\ref{sec:agn_fraction}).

We identify red sequences (RS) in 5 out of 25 galaxy groups, prominently established at $z < 1$, with early emergence in the RS locus up to $z \sim 2.2$.
The red sequence composition reveals an important finding: across the full sample, the red sequence comprises $\sim10\%$ ETGs and $\sim13\%$ LTGs of the total galaxy population.
This significant presence of LTGs on the red sequence implies low SF in disk-like galaxies, hence that environmental quenching occurs in disks before the morphological transformation to spheroidals and ellipticals. This is consistent with satellite-specific mechanisms such as ram-pressure stripping, strangulation, and tidal heating acting alongside centrally triggered feedback (Sect.~\ref{sec:red-seq}).

\cite{}Stacked $M_\star$–SFR diagnostics reveal complementary quenching pathways: starburst and main-sequence populations remain LTG dominated across all masses, while quiescent fractions become increasingly ETG dominated with stellar mass, pointing to rapid, mass-driven shutdown in the most massive galaxies and slower (perhaps environment-regulated) fading in lower-mass disks (Sect.~\ref{sec:quenching_mass_bins}).

A natural next step is a complementary wide-field study that marries COSMOS-Web with the forthcoming Euclid survey, using Euclid’s vast area and spectroscopy to extend these group diagnostics to lower-density environments, validate $R_\mathrm{cut}$ trends across the cosmic web, and test whether the AGN and quenching patterns identified here persist on the largest accessible scales.

Data Availability: Some example plots of RSD, CMDs, and SFR–Mass are shown in the Appendix, while the full set of sample plots is available at: \url{https://zenodo.org/uploads/17407954}.

\begin{acknowledgements}
      LM acknowledges the financial contribution from the
PRIN-MUR 2022 20227RNLY3 grant “The concordance cosmological model: stress-tests with galaxy clusters” supported by Next Generation EU and from the grant ASI n. 2024-10-HH.0 “Attività scientifiche per la missione Euclid – fase E.
\end{acknowledgements}

%

%

\begin{appendix} 
\section{Richest groups in successive redshift bins.}
We present a table of contents for the richest group sample, extended with the latest available spectroscopic counterparts \citep{Khostovan2025}, and complemented by a large-scale structure (LSS) analysis performed by (Taamoli, in prep.).

\begin{table*}[h!]
\caption{Richest groups in successive redshift bins.}
\label{tab:richest_clusters}
\centering
\resizebox{\textwidth}{!}{%
\begin{tabular}{|r|c|c|c|c|c|p{1.2cm}|p{1.6cm}|p{1.2cm}|c|}
\hline
ID & RA & DEC & Z & $\texttt{SN\_NO\_CLUSTER}$ & Richness & ETGs, LTGs, ALL COD$^a$ & secure, total z\textunderscore spec& $R\rm_{cut} (\mathrm{kpc})$, RSD(V-Shape) &LSS\_Flag$^b$\\
\hline
21 & 150.054 & 2.207 & 0.180 & 24.34 & 10.81 & -, +, + & 94, 144 & 400, N &-1\\
15 & 150.079 & 2.392 & 0.220 & 26.19 & 31.72 & +, +, + & 290, 364 & 700, Y & -1\\
304 & 149.784 & 2.172 & 0.340 & 13.44 & 21.46 & -, +, + & 281, 310 & 1000, N & -1\\
85 & 150.124 & 1.847 & 0.530 & 18.29 & 23.67 & +, +, + & 224, 240 & 300, N & 1\\
4 & 150.059 & 2.597 & 0.680 & 30.83 & 42.05 & -, +, + & 520, 608 & 1000, N & 2\\
1 & 149.919 & 2.517 & 0.710 & 43.11 & 77.57 & +, +, + & 429, 524 & 1000, Y & 2\\
35 & 150.084 & 2.532 & 0.890 & 21.57 & 43.51 & +, +, + & 532, 633 & 1000, Y & 2\\
36 & 150.024 & 2.202 & 0.940 & 21.26 & 41.54 & +, +, + & 647, 809 & 900, Y & 2\\
20 & 150.079 & 2.042 & 1.180 & 24.36 & 24.11 & +, +, + & 205, 293& 1000, Y & 2\\
72 & 150.194 & 2.562 & 1.310 & 18.88 & 29.59 & +, +, + & 135, 182 & 700, N & 2\\
7 & 149.994 & 2.587 & 1.560 & 29.48 & 24.75 & +, +, + & 121, 145 & 300, N & 0\\
121 & 150.099 & 2.172 & 1.770 & 16.92 & 25.54 & +, +, + & 101, 143& 400, Y & 1\\
115 & 149.984 & 2.372 & 1.880 & 17.02 & 24.39 & +, -, + & 89, 179 & 600, Y & 1\\
195 & 150.414 & 2.092 & 2.030 & 15.07 & 21.64 & -, +, + & 49, 55& 300, N & 2\\
30 & 150.299 & 1.917 & 2.220 & 22.21 & 23.58 & +, +, + & 113, 132& 100, N & 0\\
118 & 150.109 & 2.377 & 2.490 & 16.98 & 18.20 & +, -, + & 214, 310 & 500, N & 2\\
11 & 150.314 & 2.277 & 2.620 & 27.64 & 19.41 & -, +, + & 60, 113 & 300, N & 2\\
180 & 150.379 & 1.972 & 2.820 & 15.38 & 23.51 & -, +, + & 38, 61& 500, N & 0\\
82 & 150.449 & 2.457 & 2.900 & 18.32 & 22.35 & +, -, + & 36, 62& 1000, N & 2\\
156 & 150.384 & 2.502 & 3.140 & 16.01 & 27.60 & +, -, + & 25, 41& 500, N &1\\
252 & 149.879 & 2.137 & 3.260 & 14.07 & 15.12 & -, +, + & 33, 53& 200, N &0\\
380 & 149.699 & 1.987 & 3.320 & 12.59 & 15.47 & -, +, + & 21, 27& 200, N & 2\\
117 & 149.849 & 2.372 & 3.400 & 17.00 & 25.19 & +, +, + & 40, 61& 500, N & 2\\
5 & 150.294 & 1.752 & 3.610 & 30.49 & 22.82 & +, +, + & 6, 8 & 300, N & 2\\
14 & 150.319 & 2.537 & 3.650 & 26.32 & 21.56 & -, -, -  & 15, 22& 0, N & 2\\
\hline
\end{tabular}}
\\
\vskip1em 
\begin{minipage}{\textwidth}
\footnotesize{$^a$ The '+' sign indicates that the galaxy group has a central overdensity at $r < 200\,\mathrm{kpc}$ that is at least twice the density of the group's periphery.}\\
\footnotesize{$^b$ `-1' = unclassified; `0' = field; `1' = filament; `2' = cluster. }\\
\end{minipage}
\end{table*}

\section{Extended cumulative aperture results}
\label{app:cod_extended}
\begin{figure}
    \centering
    \begin{subfigure}{\columnwidth}
        \includegraphics[width=\columnwidth]{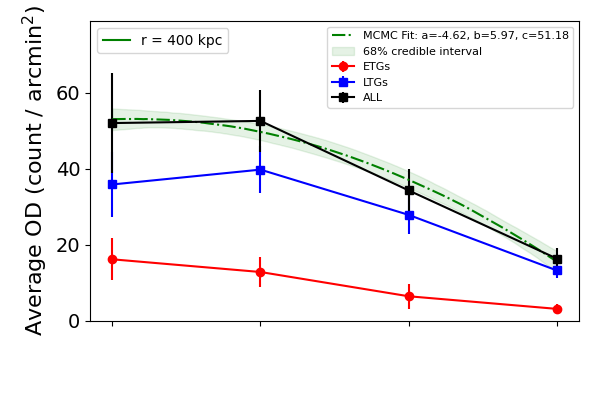}
    \end{subfigure}

    \vspace{-1.5em}

    \begin{subfigure}{\columnwidth}
        \includegraphics[width=\columnwidth]{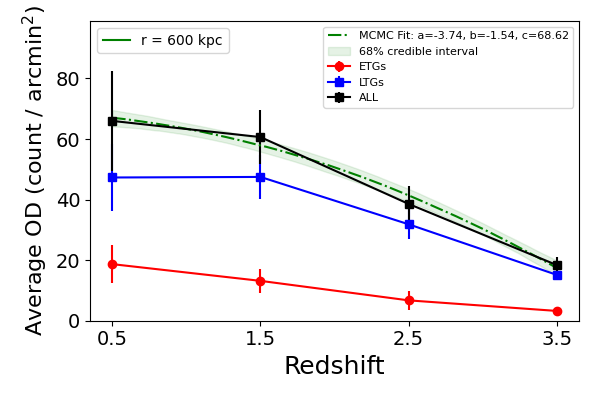}
    \end{subfigure}
    \caption{
    Average overdensity versus redshift for ETGs (red circles) and LTGs (blue squares), measured within cumulative apertures of 400 and $600\,\mathrm{kpc}$. The MCMC fits and their 68\% credible intervals are shown as dashed curves and shaded regions. These panels extend the results shown in Figure~\ref{fig:cum-cod_200} for the $200\,\mathrm{kpc}$ aperture.}
    \label{fig:cum-cod-200-400}
\end{figure}

Extended cumulative overdensity profiles at 400 and $600\,\mathrm{kpc}$ are shown in Fig.~\ref{fig:cum-cod-200-400}, consistent with the $200\,\mathrm{kpc}$ trends: ETG densities decline while LTG densities slightly rise, highlighting persistent star-forming disks in group centers.

\section{Photometry Completeness}
\label{sec:phot_comp}

\begin{figure}
    \centering
        \includegraphics[width=0.49\textwidth]{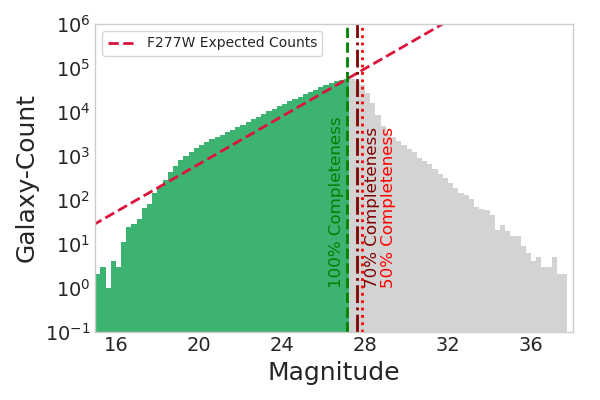}
    \caption{F277W completeness diagnostics showing galaxy counts as a function of total magnitude. Light-green and gray histograms trace the observed number counts, the dashed red line highlights the linear fit to the bright end, and vertical dashed markers mark the 100\% (green), 70\% (maroon), and 50\% (red) completeness limits adopted in this work.}
    \label{fig:multi_filter_completeness}
\end{figure}

We evaluate photometric completeness by fitting the bright-end number counts and measuring deviations as a function of magnitude, identifying the 100\% level at the histogram peak. Table~\ref{tab:completeness} and Figure~\ref{fig:multi_filter_completeness} summarize the resulting limits for F814W, F277W, and F444W, and we adopt a conservative 95\% completeness threshold for the remainder of the analysis.
\begin{table}[htbp]
\centering
\caption{Photometric completeness limits for F814W, F277W, and F444W.}
\label{tab:completeness}
\begin{tabular}{lccc}
\toprule
Filter & \multicolumn{1}{p{1.5cm}}{\centering Mag at 100\% Comp} &
         \multicolumn{1}{p{1.5cm}}{\centering Mag at 70\% Comp} &
         \multicolumn{1}{p{1.5cm}}{\centering Mag at 50\% Comp} \\
\midrule
F814W & 27.62 & 28.12 & 28.38 \\
F277W & 27.12 & 27.62 & 27.88 \\
F444W & 27.38 & 27.88 & 28.12 \\
\bottomrule
\end{tabular}

\vspace{0.5em}  
{\footnotesize We adopt a conservative 95\% completeness threshold for the analysis.}

\end{table}

\section{Mass completeness}\label{sec:stella_mass_completeness}
We calculate stellar mass completeness using the \citet{Pozzetti2010} method:
\begin{equation}
\label{eq:stellar_mass_completeness}
\log_{10}(M_{\text{resc}}) = \log_{10}(M_\star/\modot) + 0.4(m_{\text{F444W}} - m_{\text{lim}})
\end{equation}

This involves deriving a rescaled stellar mass ($M_{\text{resc}}$) for each galaxy by scaling its stellar mass to the survey's limiting magnitude. After performing this rescaling for the entire sample, the limiting stellar mass ($M_{\text{lim}\star}$) is defined as the 90th percentile of the resulting $M_{resc}$ distribution. This process provides a reliable indicator of the mass down to which the galaxy sample is statistically complete.
We have an excellent sample of galaxies at $z \sim 4$, with complete samples down to $\log_{10}M_{\star} \sim 8.5\,\modot$ at the highest redshifts, and down to $\log_{10}M_{\star} \sim 7\,\modot$ in the nearby Universe, having a mass
completeness improvement by about 1 dex compared to COSMOS2020 \citep{Shuntov2025}.

\begin{figure}[ht]
    \centering
    \includegraphics[width=0.45\textwidth]{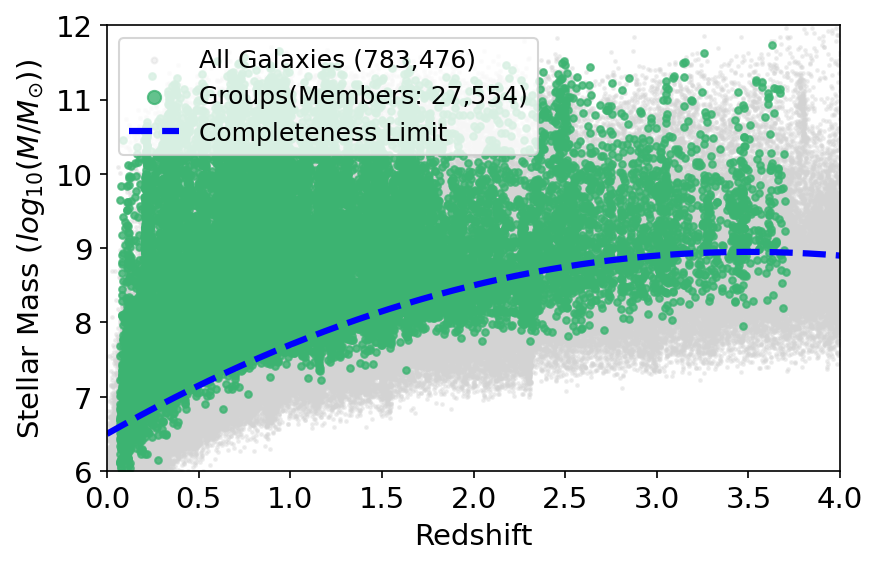}
    \caption{Stellar mass as a function of redshift for all catalog galaxies (gray) and confirmed group members (green). The dashed blue curve marks the stellar-mass completeness limit adopted in this analysis.}
    \label{fig:mass_completeness}
\end{figure}
\section{Group radial surface density and annuli plots}
We show a selection of radial density and annuli  plots, similar to Fig.~\ref{fig:hist-rsd-annuli}, in here. The remaining plots are available online via the Zenodo link: \url{https://zenodo.org/uploads/17407954}.

\begin{figure*}[htbp]
    \centering
    \includegraphics[width=0.98\textwidth, keepaspectratio]{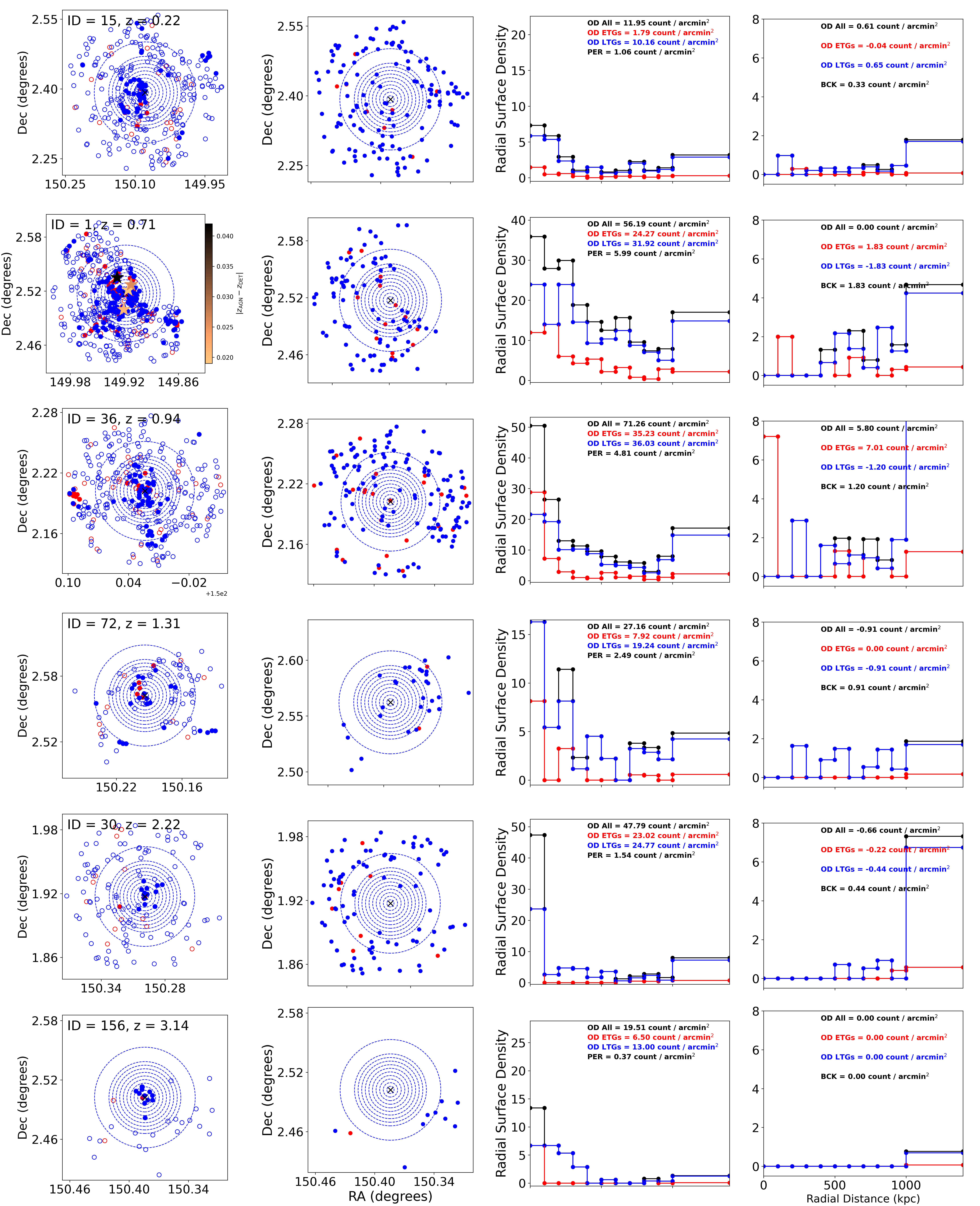}
    \caption{Annuli and radial surface-density diagnostics for group and field galaxy candidates; subsequent panels repeat the layout for additional groups.}
    \label{fig:combined_plot_rsd}
\end{figure*}

\section{Customized red sequence model parameters}
We summarize the key parameterizations used in our red sequence galaxy model. These include the redshift evolution and environmental metallicity boosts, the distribution of galaxy formation redshifts, dust attenuation, and the star formation history timescales. Each subsection contains the numerical prescriptions and scaling relations adopted in the model, allowing readers to trace how stellar mass, redshift, and environment influence the predicted properties of red sequence galaxies.

\subsection{Redshift evolution and environmental metallicity boosts}\label{sec:redshift_evolution_and_metallicity_boost}

The mass-metallicity relation (MZR) forms the foundation of our red sequence model. We extend the Sanders et al. (2021) MZR to include redshift evolution and environmental effects, while maintaining consistency with local observations \citep{Tremonti2004, Gallazzi2005}. The redshift evolution term $\Delta_z$ captures the systematic decrease of metallicity with increasing redshift, reflecting both high-redshift galaxy observations and theoretical expectations for chemical enrichment across cosmic time.

The redshift evolution parameters adopted are defined as follows:
\begin{itemize}
    \item $z \le 1.0$: $\Delta_z = -0.14z$
    \item $1<z \le 2.5$: $\Delta_z = -0.14-0.20(z-1.0)$
    \item $z >2.5$: $\Delta_z = -0.44-0.35(z-2.5)$
\end{itemize}

The environmental metallicity boost accounts for the observation that galaxies in groups and clusters are more metal-rich than field galaxies at fixed stellar mass \citep{Pasquali2012, Peng2015}. This enhancement depends on stellar mass and decreases with redshift, reflecting the reduced efficiency of environmental processing in the early Universe \citep{Fossati2017, Chartab2020}.

The environmental metallicity boost parameters adopted are defined as follows:
\begin{itemize}
    \item Baseline enhancement: 0.07\,dex at low redshift
    \item Mass-dependent term (in solar masses): 0.02\,dex per dex above $\log_{10} M_\star = 10.5$
    \item Redshift suppression: decreases with $z$, roughly $\exp(-z/2)$
    \item High-redshift reduction: for $z>3$, both terms are multiplied by 0.3
\end{itemize}

\subsection{Formation redshift distribution}
\label{sec:formation_redshift}
ZGH{Formation redshift is the redshift at which a galaxy formed 50\% of its stellar mass}. Galaxy formation redshifts are assigned based on stellar mass and observation redshift following downsizing trends \citep{Thomas2010, McDermid2015, Behroozi2013, Rodriguez-Gomez2016, Nakajima2023, Curti2024}. The downsizing paradigm predicts that more massive galaxies form earlier and more rapidly than lower-mass systems. This approach ensures that our synthetic red sequence galaxies have realistic formation histories that match observed trends.

The assigned formation redshift is limited to $z_{\rm obs}+2$ and capped at 10, with formation times shorter than $0.1\,\mathrm{Gyr}$ avoided to ensure physically realistic evolution. The mass-dependent formation redshifts reflect the observed correlation between galaxy mass and stellar age.

The formation redshift assignments by stellar mass adopted are defined as follows:
\begin{itemize}
    \item $\log_{10}(M_\star/\modot)\geq11.5$: $z_{\rm form} = 4,5,6$ for $z_{\rm obs}=1,2,3$; up to $\min(8.0, z_{\rm obs}+1.5)$
    \item $10.5 \leq \log_{10}(M_\star/\modot) < 11.5$: $z_{\rm form} = 3,4,5$ for $z_{\rm obs}=1,2,3$; up to $\min(6.5, z_{\rm obs}+1.0)$
    \item $9.5 \leq \log_{10}(M_\star/\modot) < 10.5$: $z_{\rm form} = 2,3$ for $z_{\rm obs}=1,2$; up to $\min(4.5, z_{\rm obs}+0.8)$
    \item $\log_{10}(M_\star/\modot)<9.5$: $z_{\rm form} = 1.5,2.5$ for $z_{\rm obs}=1,2$; up to $\min(3.5, z_{\rm obs}+0.5)$
\end{itemize}

\subsection{Dust attenuation for red sequence galaxies}
\label{sec:dust}

Dust attenuation for red sequence galaxies is modeled as a function of stellar mass and redshift, following observational constraints \citep{Whitaker2017, Bouwens2016, Fudamoto2021}. More massive galaxies exhibit slightly higher $V$-band attenuation, while attenuation increases moderately with redshift. To account for environmental effects, a small reduction is applied for galaxies in dense regions, reflecting processes such as gas stripping that can remove dust from group environments.

The resulting prescription ensures low but evolving dust levels, with an upper limit of $A_V = 0.3\,\mathrm{mag}$, consistent with the typically quiescent nature of red sequence systems. This approach captures the observed trend that red sequence galaxies have lower dust content than star-forming systems while still accounting for the mass and redshift dependencies observed in quiescent populations.
The dust attenuation parameters adopted are defined as follows:
\begin{itemize}
    \item Mass contribution: $0.08\max(0, \log M_\star - 10.0)$
    \item Redshift contribution:
    \begin{itemize}
        \item $z < 1$: $0.03z$
        \item $1\leq z<2.5$: $0.03+0.05(z-1)$
        \item $z \geq 2.5$: $0.13-0.08(z-2.5)$
    \end{itemize}
    \item Environmental reduction: $-0.04(1+z)^{-0.5}$
\end{itemize}

\subsection{Star formation history timescale ($\tau$)}
\label{sec:sfh_timescale}

The age spread of red sequence galaxies is modeled through exponentially declining star formation histories characterized by a timescale $\tau$, which depends on stellar mass and redshift \citep{Thomas2010, McDermid2015}. More massive galaxies are assigned shorter $\tau$, reflecting their rapid, early formation, while lower-mass systems have more extended star formation histories. The timescale also evolves mildly with redshift, allowing slightly longer star formation at earlier epochs.

A minimum value of $\tau = 0.05\,\mathrm{Gyr}$ is imposed to ensure physically realistic evolution. This prescription captures the observed downsizing trend and provides realistic age spreads for synthetic red sequence galaxies, ensuring that our model produces galaxies with formation histories consistent with observed quiescent populations.

The star formation history timescale parameters adopted are defined as follows:
\begin{itemize}
    \item Base $\tau$ ($\mathrm{Gyr}$) by stellar mass:
    \begin{itemize}
        \item $\log M_\star \geq 11.5$: $\tau_{\rm base} = 0.05$
        \item $10.5 \leq \log M_\star < 11.5$: $\tau_{\rm base} = 0.15$
        \item $9.5 \leq \log M_\star< 10.5$: $\tau_{\rm base} = 0.3$
        \item $\log M_\star < 9.5$: $\tau_{\rm base} = 0.6$
    \end{itemize}
    \item Redshift scaling:
    \begin{itemize}
        \item $z < 1.0$: $\tau = \tau_{\rm base} (1 + 0.1z)$
        \item $1.0 \leq z < 2.5$: $\tau = \tau_{\rm base} (1.1 + 0.05 (z - 1.0))$
        \item $z \geq 2.5$: $\tau = \tau_{\rm base} (1.15 - 0.1 (z - 2.5))$
    \end{itemize}
\end{itemize}
%
\section{SFR--Mass, CCD, and CMD plots}
\begin{figure*}[htbp]
    \centering
    \includegraphics[width=0.98\textwidth]{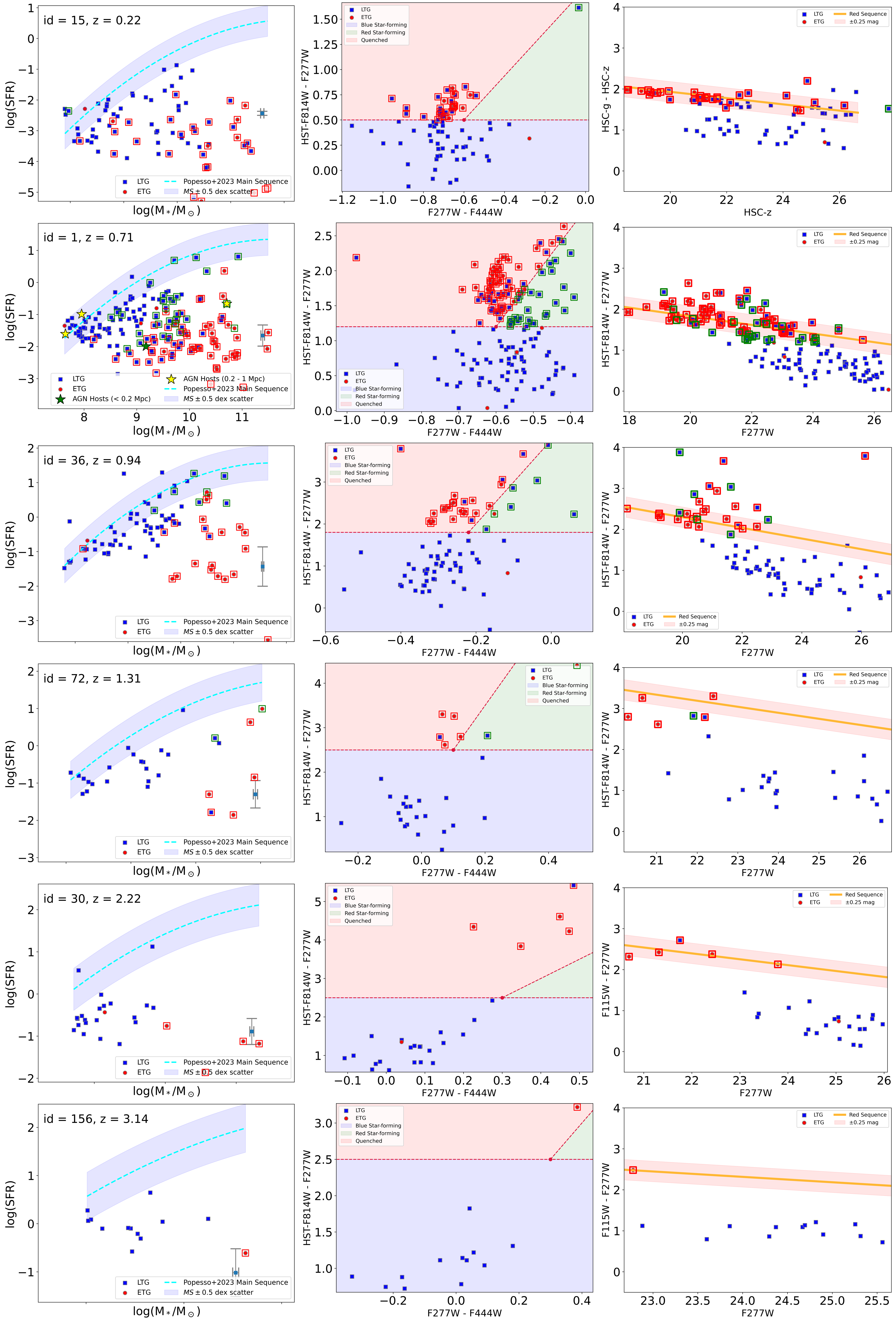}
    \caption{As Fig.~\ref{fig:combined_plots_sfr_ccd_cmd} but for additional galaxy groups.}
    \label{fig:combined_plot_sfr} 
\end{figure*}

We show a selection of SFR–Mass, CCD, and CMD plots, similar to Fig.~\ref{fig:combined_plot_sfr}, in here. The remaining plots are available online via the Zenodo link: \url{https://zenodo.org/uploads/17407954}.

\end{appendix}
\end{document}